\documentclass[pra,twocolumn,showpacs,amsmath,amssymb,superscriptaddress,longbibliography,floatfix]{revtex4-1}
\usepackage[none]{hyphenat}
\usepackage{graphicx}
\usepackage{amsmath, amsfonts, amssymb}
\usepackage{mathtools}
\usepackage{physics}
\usepackage[commandnameprefix=always]{changes}
\usepackage{hyperref}

\definecolor{myblue}{rgb}{0.18039,0.1882353,0.57255}
\definecolor{myred}{rgb}{1,0.,0.3}
\definechangesauthor[name={Per cusse}, color=red]{per} 
\hypersetup{
    colorlinks=true,
    citecolor=myblue,
    linkcolor=myblue,
    urlcolor=myblue,
}
\usepackage{orcidlink}
\begin{document}
\preprint{APS/123-QED}

\title{Exotic Collective Behaviors of  Giant Quantum Emitters in High-Dimensional Baths}

\author{Qing-Yang Qiu\orcid{0009-0007-3214-4892}}
\affiliation{School of Physics and Institute for Quantum Science and Engineering, Huazhong University of Science and Technology, Wuhan, 430074, China}
\affiliation{Wuhan institute of quantum technology, Wuhan, 430074, China}

\author{Wen Huang}
\affiliation{School of Physics and Institute for Quantum Science and Engineering, Huazhong University of Science and Technology, Wuhan, 430074, China}
\affiliation{Wuhan institute of quantum technology, Wuhan, 430074, China}

\author{Lei Du}
\affiliation{Department of Microtechnology and Nanoscience, Chalmers University of Technology, Gothenburg, 41296, Sweden}

\author{Xin-You L\"{u}}\email{xinyoulu@hust.edu.cn}
\affiliation{School of Physics and Institute for Quantum Science and Engineering, Huazhong University of Science and Technology, Wuhan, 430074, China}
\affiliation{Wuhan institute of quantum technology, Wuhan, 430074, China}

\date{\today}
\begin{abstract}
Nonlocal light-matter interactions with giant atoms in high-dimensional environments are not only fundamentally intriguing for testing quantum electrodynamics beyond the dipole approximation but also crucial for building high-dimensional quantum networks and engineering multipartite entangled states. Given the enigmatic and largely uncharted collective signatures exhibited by multiple giant atoms within high-dimensional optical baths, we delve into their non-perturbative collective dynamics when coupled to a common high-dimensional photonic reservoir, employing a resolvent operator approach. We demonstrate that precisely engineered atomic arrangements lead to unconventional quantum dynamics, featuring non-Markovianity-induced beats and long-lived bound states in the continuum, thereby providing a versatile platform for implementing high-dimensional quantum memory. Phenomenologically, we observe the emergence of exotic photon emission patterns in both 2D and 3D baths. The emission directions are shown to be precisely controllable on demand through exact phase engineering of the coupling parameters, enabling a highly efficient chiral light-matter interface. Moreover, our generalization to a 3D bath reveals that coherent dipole-dipole interactions can survive despite the coupling to a continuum of modes, a finding that challenges conventional wisdom regarding decoherence.
\end{abstract}

\maketitle
\section{INTRODUCTION}\label{Sec I}
Photon emission lies at the heart of light-matter interactions and serves as a cornerstone of photonic quantum science\,\cite{GROSS1982301,PhysRevLett.76.2049,Eschner2001}. This fundamental process was first examined in depth by Dicke in his seminal work\,\cite{Dicke1954}, showing that emitters without direct dipole-dipole interactions can undergo collective decay via shared bath modes. Such emergent quantum cooperativity critically depends on the structured bath spectral densities with one\,\cite{PhysRevA.93.033833} or multiple energy bands\,\cite{PhysRevA.97.043831,PhysRevLett.126.203601,PhysRevA.105.023703,PhysRevB.108.045407}, nonlinearities\,\cite{PhysRevLett.124.213601,PRXQuantum.4.030326}, band topology\,\cite{sciadv0297,PhysRevLett.124.023603}, spatial dimensionality\,\cite{PhysRevLett.132.163602,PhysRevLett.128.113601}, and inter-emitter delays\,\cite{PhysRevLett.124.043603}.

Giant atoms (GAs), characterized by their exotic self-interference effects and nonlocal coupling nature\,\cite{PhysRevA.90.013837,PhysRevA.95.053821,Frisk_Kockum_2020}, have emerged as a rapidly growing research frontier in waveguide quantum electrodynamics (QED). Leveraging the state-of-the-art waveguide QED platforms utilizing surface acoustic waves\,\cite{Andersson2015} or meandering coplanar waveguide\,\cite{Kannan2020,PhysRevA.103.023710,PhysRevX.13.021039}, this exotic atomic geometry has been successfully implemented in one-dimensional (1D) architectures, with recent extension to ``giant" ferromagnetic spin ensemble\,\cite{Wang2022}.  A rich variety of phenomena that are absent in natural atomic systems have been uncovered, including decoherence-free interactions\,\cite{PhysRevLett.120.140404,PhysRevA.107.023705,PhysRevResearch.2.043184}, oscillating\,\cite{PhysRevResearch.2.043014,PhysRevA.106.013702,PhysRevA.107.023716,Xu_2024} or tunable bound states\,\cite{PhysRevLett.126.043602}, robust generation of entanglement\,\cite{PhysRevLett.130.053601,PhysRevA.108.023728,PhysRevA.111.033707}, and non-exponential decay\,\cite{Qiu2023,PhysRevResearch.6.033243,Qiulaser}. Despite these advances, the generalization of nonlocal light-matter interactions to high-dimensional reservoirs\,\cite{PhysRevLett.122.203603,Redondo2021,PhysRevResearch.6.043222,Leonforte_2025}, particularly regarding collective signatures, is undoubtedly of paramount importance, given their potential to revolutionize high-dimensional quantum networks and unravel exotic many-body quantum phases\,\cite{PhysRevResearch.3.043203,GONG2023906,PhysRevA.108.033314}.

In this manuscript, we employ versatile resolvent techniques\,\cite{Cohen_book} to provide a unified and non-perturbative framework for studying geometry-dependent quantum dynamics of GAs in both 2D and 3D waveguide QED. While previous studies have explored GAs in structured photonic environments, our work delivers several key advances. First, we present the fully analytical treatment of multi-GA dynamics in a 2D square lattice, explicitly connecting atomic geometry to non-Markovian decay rates and bound state formation. When the atomic transition frequencies lie close to the 2D Van Hove singularity, the dynamics of SGAs exhibits pronounced beats combined with the enhanced non-Markovian information backflow. In contrast, DGAs support a long-lived bound state in the continuum (BIC), enabling high-dimensional quantum information storage. Furthermore, we also report exotic radiation patterns with no prior analogue in lower dimensions. These include multidirectional emission with spatially condensed energy density and quasi-1D radiative channels, both arising from tailored interference within the high-dimensional reservoir. Notably, we demonstrate that by imprinting complex phases on the couplings, GAs can serve as chiral quantum sources, enabling deterministic photon routing--a functionality crucial for quantum networks.
Most importantly, our work constitutes the first theoretical extension of GAs electrodynamics to 3D structured reservoirs.
\begin{figure}
  \centering
  \includegraphics[width=8.7cm]{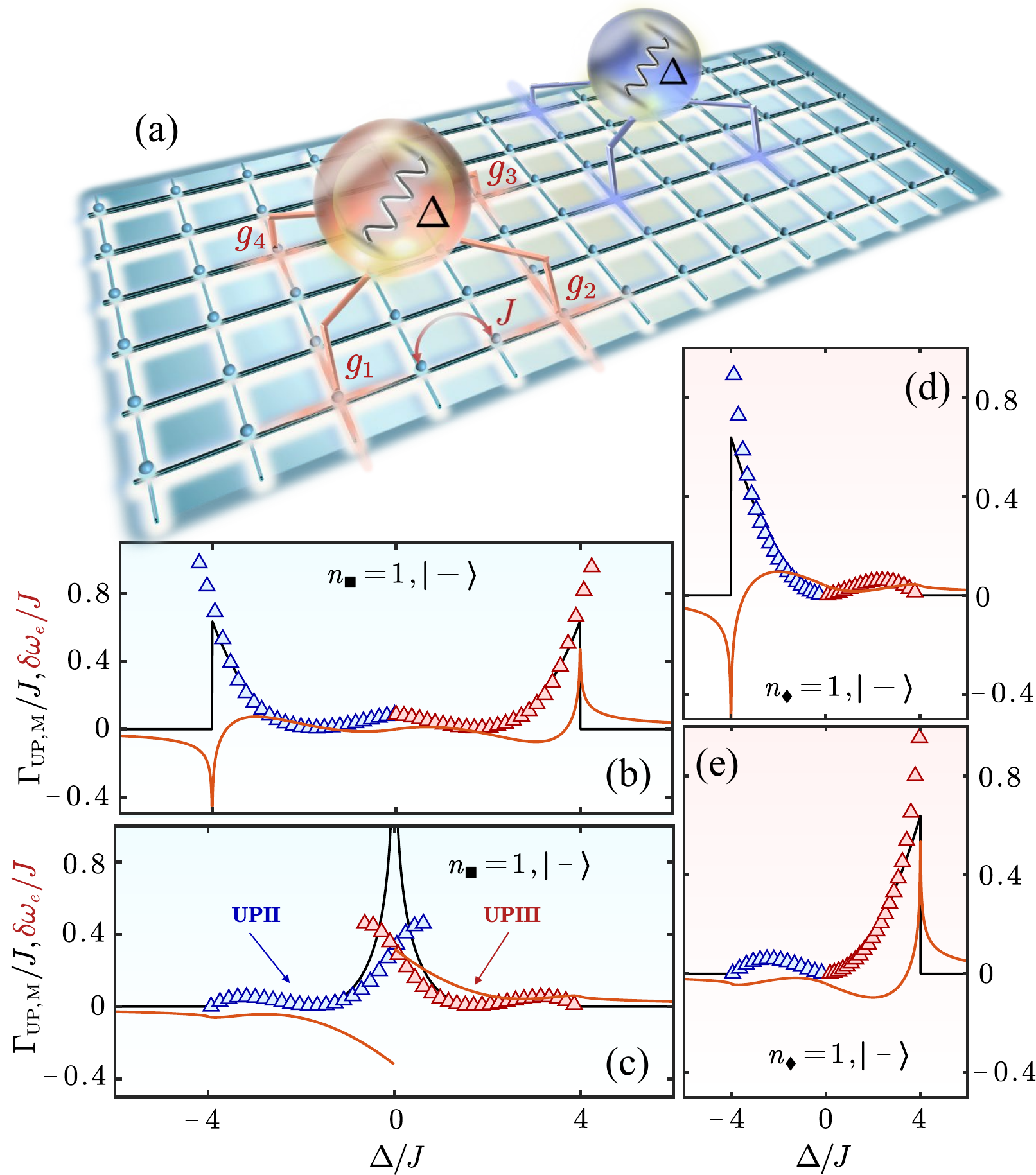}
  \caption{(a) Schematic representation: A pair of two-level GAs coupled to a photonic lattice with nearest-neighbor hopping $J$. We plot the Markovian (black solid lines) and non-Markovian (triangles) decay rates, along with the corresponding frequency shift $\delta\omega_{e}$ (red solid lines), as a function of detuning $\Delta$. The simulations in panels (b-c) [(d-e)] are performed by considering two coupled SGAs (DGAs), with the physical parameters specified in the legend. All panels share the identical coupling $g=0.2J$.}\label{fig1}
  \vspace{-10pt} 
\end{figure}
In this setting, we predict exotic emission profiles and geometry-dependent decoherence-free interactions, supported by analytical solutions. These results establish an innovative paradigm for high-dimensional quantum optics and offer design principles for scalable multi-qubit architectures.  All these predicted phenomena originate from geometry-controlled interference networks within high-dimensional electromagnetic fields. Our work establishes a paradigm for engineering light-matter interactions beyond low-dimensional limits, laying the groundwork for scalable quantum networks.

The manuscript is organized as follows. In Sec.\ref{Sec II}, we describe the system under study by presenting the full Hamiltonian and detailing the two types of atomic coupling configurations considered. In Sec.\ref{Sec III}, we characterize the non-Markovian behavior of a pair of GAs through their exact decay rates. Section \ref{Sec IV} is devoted to the collective dynamics of both atomic and photonic subsystems, where exotic emission patterns emerge due to delicate interference effects between the GAs. In Sec.\ref{Sec V}, we explore chiral emission by introducing nontrivial phases in the nonlocal light–matter couplings. In Sec.\ref{Sec VI}, we investigate decoherence-free interactions in a pair of diamond-type GAs, extending our results to a three-dimensional configuration. Finally, in Sec.\ref{Sec VII}, we discuss potential experimental implementations and conclude by summarizing the main findings and suggesting future research directions.

\section{SYSTEM}\label{Sec II}
We are interested in the collective light-matter interactions of GAs bathed in a 2D structured environment.  As depicted in Fig.\,\ref{fig1}(a), the system is composed of a pair of GAs interacting with a 2D photonic lattice through multiple coupling points. The bath can be modeled by a square lattice of $N\times N$ bosonic modes characterized by identical energies $\omega_{c}$ and annihilation operators $a_{\boldsymbol{n}}\equiv a_{n_{x},n_{y}}$, featuring exclusively nearest-neighbor coupling $J$. Despite its simplicity, this system captures most of the prominent features of experimentally available models\,\cite{PhysRevLett.104.203603,PhysRevLett.106.096801,PhysRevX.2.011014,science.1237125}. The total Hamiltonian of the GAs plus bath reads $H_{{\rm{tot}}}=H_{A}+H_{B}+H_{{\rm{int}}}$, where $H_{A}=\sum_{\ell=1,2}\Delta\sigma_{\ell}^{\dagger}\sigma_{\ell}^{}$ and $H_{B}=-J\sum_{\langle \boldsymbol{m}, \boldsymbol{n}\rangle}(a_{\boldsymbol{n}}^{\dagger}a_{\boldsymbol{m}}^{}+{\rm{H.c.}})$ are the free Hamiltonians for the GAs and the field, respectively, with atomic coherent operators $\sigma_{\ell}\equiv\ket{g}_{\ell}\bra{e}$ and detuning $\Delta\equiv\omega_{e}-\omega_{c}$. Note that we have chosen a rotating frame at $\omega_{c}$ such that the band center is energetically pinned to zero.  Under the rotating wave approximation, the emitter-field interaction is described by a nonlocal Hamiltonian $H_{{\rm{int}}}=\sum_{\ell=1,2}\sum_{p=1}^{M}g_{p}(a_{\boldsymbol{n}_{\ell p}}^{\dagger}\sigma_{\ell}^{}+{\rm H.c.})$, where $\boldsymbol{n}_{\ell p}$ represents the location of the bosonic mode that interacts with the $p\,$-th coupling point of the $\ell$-th atom, at which the coupling strength is given by $g_{p}$. Unless otherwise specified, we adopt uniform couplings with $g_{p}=g$ in our analysis. Diagonalizing bath Hamiltonian in reciprocal space yields $H_{B}=\sum_{\boldsymbol{k}}\omega(\boldsymbol{k})a^{\dagger}_{\boldsymbol{k}}a_{\boldsymbol{k}}^{}$ with dispersion relation $\omega(\boldsymbol{k})=$ $-2J[\cos(k_{x})+\cos(k_{y})]$ and discrete wave numbers $k_{x},k_{y}\in \frac{2\pi}{N}(-\frac{N}{2},\cdots,\frac{N}{2}-1)$. Here, periodic boundary conditions are imposed, and the real-space operators obey $a_{\boldsymbol{n}}=\frac{1}{N}\sum_{\boldsymbol{k}} a_{\boldsymbol{k}}e^{-i\boldsymbol{k}\cdot\boldsymbol{n}}$.

For a pair of GAs, it is instructive to rewrite the interaction Hamiltonian as
\begin{align}\label{eq1}
H_{{\rm int}}=\frac{\sqrt{2}g}{N}\sum_{\boldsymbol{k}\in\boldsymbol{k}^{{\rm I}}}\sum_{\alpha=\pm^{}}\left[\sqrt{\mathcal{P}_{\alpha}(\boldsymbol{k})}a_{\boldsymbol{k},\alpha}^{\dagger}\sigma^{}_{\alpha}+{\rm H.c.}\right],
\end{align}
where the summation over momentum $\boldsymbol{k}^{{\rm I}}$ is restricted to the first quadrant $\boldsymbol{k}^{{\rm I}}\equiv\{\boldsymbol{k}| k_{x},k_{y}>0\}$ and $\sigma_{\pm}\equiv(\sigma_{1}\pm\sigma_{2})/\sqrt{2}$. The explicit expressions for the modified bath modes $a_{\boldsymbol{k},\pm}$ and the constants $\mathcal{P}_{\alpha}(\boldsymbol{k})$ are provided in Appendix \ref{A}. It turns out that these modes are mutually orthogonal and satisfy bosonic commutation relation $[a^{}_{\boldsymbol{k},\alpha},a^{\dagger}_{\boldsymbol{k}',\alpha'}]\!=\!\delta_{\boldsymbol{k},\boldsymbol{k}'}\delta_{\alpha\alpha'}$. Therefore, the dynamics of collective states $\ket{\pm}\equiv\sigma^{\dagger}_{\pm}|gg,\{0\}\rangle$ become decoupled, allowing them to be treated independently. Here, $|gg,\{0\}\rangle$ is the total ground state, where the atoms are in their lower state $\ket{g}$ and the waveguide is empty.

\begin{figure}
  \centering
  \includegraphics[width=8.5cm]{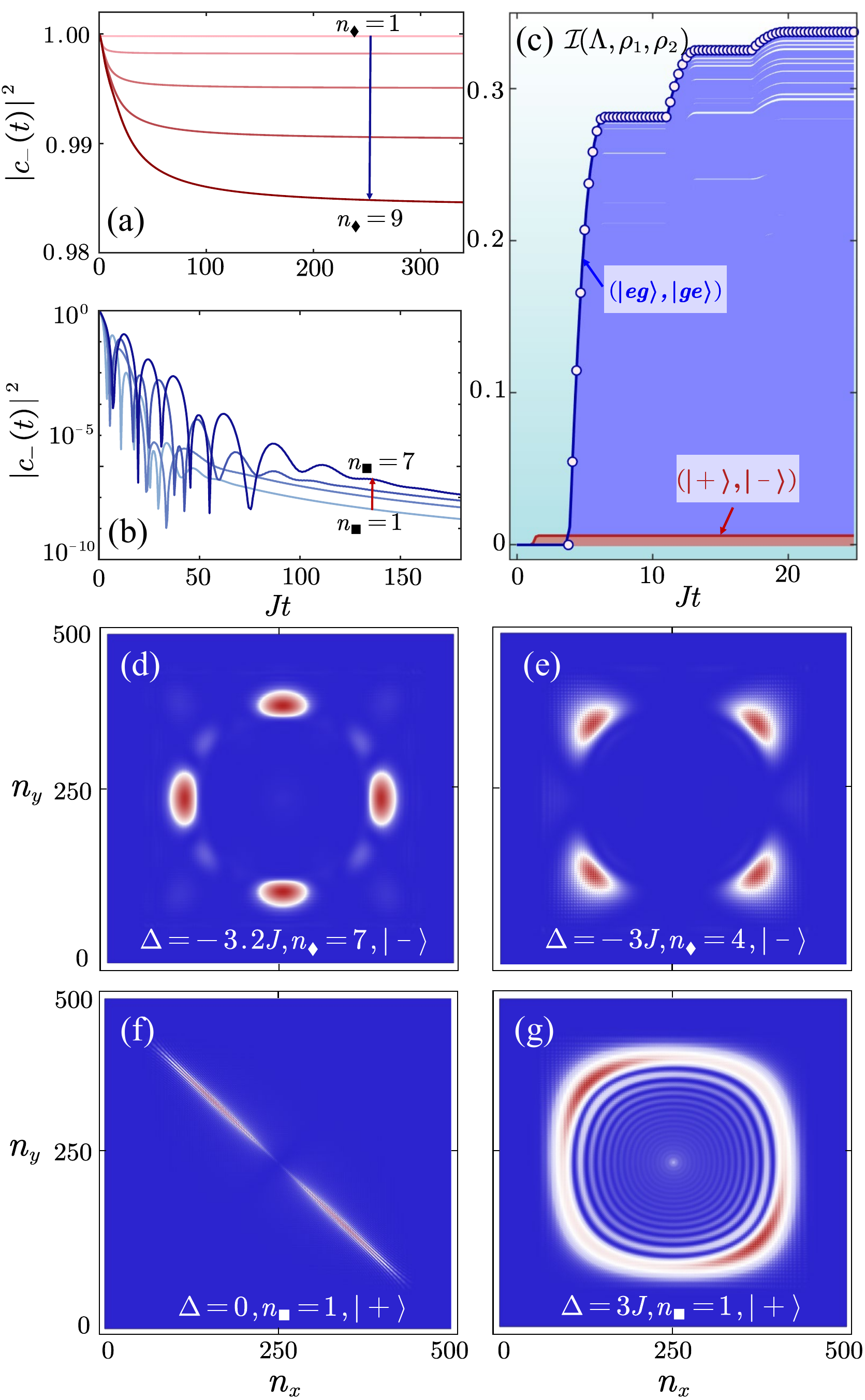}
  \caption{(a)-(b) The antisymmetric population $|c_{-}(t)|^{2}$ for a pair of DGAs (SGAs) with different atomic size $n_{\blacklozenge}=1,3,5,7,9\, (n_{\protect\scalebox{0.55}{$\protect\blacksquare$}}=1,3,5,7)$. The latter is plotted in logarithmic scale.  Panel (c) plots the integrated non-Markovian measure $\mathcal{I}(\Lambda,\rho_{1},\rho_{2})$ as a function of $Jt$ for SGAs with $n_{\protect\scalebox{0.55}{$\protect\blacksquare$}}\!=\!1$ by performing time evolution initiated from 10000 random initial state pairs $(\rho_{1},\rho_{2})$. The evolutionary trajectories maximizing the non-Markovianity $\mathcal{N}(\Lambda)$ are explicitly identified. Panels (a)-(c) share the same detuning $\Delta=0$, while the coupling strength is set to $g=0.01J$ in panel (a) and $g=0.25J$ in panels (b) and (c). We also plot the bath population $|C_{\boldsymbol{n}}^{\pm}(t)|$ in real space [(d)-(g)] at time $tJ=100$ with $g=0.2J$, where the performed physical parameters are speciﬁed in the legend.}\label{fig2}
  \vspace{-10pt} 
\end{figure}

In this work, we investigate two distinct coupling configurations of GAs: square-like GAs (SGAs) and diamond-like GAs (DGAs). In the SGA configuration, one emitter is centered at the origin and couples to the waveguide at four symmetric positions $(\pm n_{\protect\scalebox{0.55}{$\protect\blacksquare$}}, \pm n_{\protect\scalebox{0.55}{$\protect\blacksquare$}})$ and $(\pm n_{\protect\scalebox{0.55}{$\protect\blacksquare$}}, \mp n_{\protect\scalebox{0.55}{$\protect\blacksquare$}})$. The other emitter has an identical coupling pattern but is displaced from the origin by a relative position vector $\boldsymbol{n}_{c} = (n_{\protect\scalebox{0.55}{$\protect\blacksquare$}}, n_{\protect\scalebox{0.55}{$\protect\blacksquare$}})$. In the DGA configuration, one emitter couples at four locations $(\pm n_{\blacklozenge}, 0)$ and $(0, \mp n_{\blacklozenge})$, also centered symmetrically around the origin. Similarly, the other emitter follows the same spatial coupling profile, shifted by a vector $\boldsymbol{n}_{c} = (n_{\blacklozenge}, 0)$. It is worth emphasizing that these geometric arrangements are inspired by the braided structures initially proposed in 1D architectures\,\cite{PhysRevLett.120.140404}.

\section{NON-MARKOVIAN COLLECTIVE DECAY}\label{Sec III}
We now characterize the collective properties of emitters in the band regime through their decay rates. The exact decay rates of states $\ket{\pm}$ are quantified by solving pole equations $z-\Delta-\Sigma_{\pm}(z)=0$. Here, $\Sigma_{\pm}(z)$ are the self-energies incorporating the bath-induced corrections to collective dynamics of emitter states $\ket{\pm}$. For a pair of GAs with identical spatial configurations, they take a simple form of $\Sigma_{\pm}(z)=\Sigma_{e}(z)\pm\Sigma_{12}(z)$, where the definitions of individual and collective components are displayed in Appendix \ref{A}. Given a spatial configuration of emitters, the self-energies can be determined analytically. For instance, when considering a pair of DGAs with $n_{\blacklozenge}=1$, the self-energies exhibit simple analytical forms (see Appendix \ref{A} for more details)
\begin{align}\label{eq2}
\small
\!\!\!\Sigma_{\pm}(z)=\frac{g^{2}}{4J}\left(\frac{4z}{J}\mp\frac{z^{2}}{J^{2}}\right)\!\left(\frac{2}{\pi }K\!\left[\left(\frac{4J}{z}\right)^{2}\right]\!-\!1\!\right)\pm\frac{g^{2}}{J},
\end{align}
where $K(\bullet)$ is the complete elliptical integral of the first kind. Specially, close to the band center (for $\varepsilon/J\ll 1$), the symmetric self-energy $\Sigma_{+}(\varepsilon\!+\!i0^{+})$ in Eq.\,(\ref{eq2}) can be expanded as
\begin{align}\label{eq3}
\small
\Sigma_{+}(\varepsilon+i0^{+}) \approx \frac{g^{2}}{J}-i\frac{g^{2}\varepsilon^{2}}{4\pi J^{3}} \ln\left[\left(\frac{16J}{\varepsilon}\right)^{2}\right],
\end{align}
which perturbatively captures the vanishing decay of states at $\Delta = 0$, i.e., $\left.{\rm{Im}}[\Sigma_{+}(\varepsilon+i0^{+})]\right|_{\varepsilon\rightarrow 0}=0$, while the exact dynamics requires more rigorous non-perturbative treatments.

In Figs.\,\ref{fig1}(b-e), we compare the Markovian and non-Markovian decay rates for two types of GAs with $n_{\blacklozenge/\protect\scalebox{0.55}{$\protect\blacksquare$}}=1$. Here, the Markovian decay rates $\Gamma_{{\rm M}}$ are determined analogously, but replacing $\Sigma_{\alpha}(z)\!\rightarrow\!\Sigma_{\alpha}(\Delta+i0^{+})$ in the pole equation. We denote the exact unstable poles as $\text{UPII}$ or $\text{UPIII}$ when their real parts lie in $[-4J,0)$ or $(0,4J]$,  respectively. To sum up, these decay rates exhibit symmetric (for SGAs) and asymmetric (for DGAs) profiles about the band center. For SGAs, the exact decay rates of state $\ket{+}/\ket{-}$ demonstrate enhanced agreement with Markovian predictions near the band center/edges, respectively; while for DGAs, the agreement improves near the upper/lower band edges for state $\ket{+}/\ket{-}$. More detailed analyses, including those for larger atomic systems, are provided in Appendix \ref{B}, offering comprehensive insights into the relaxation dynamics.

\begin{figure}
  \centering
  \includegraphics[width=8.7cm]{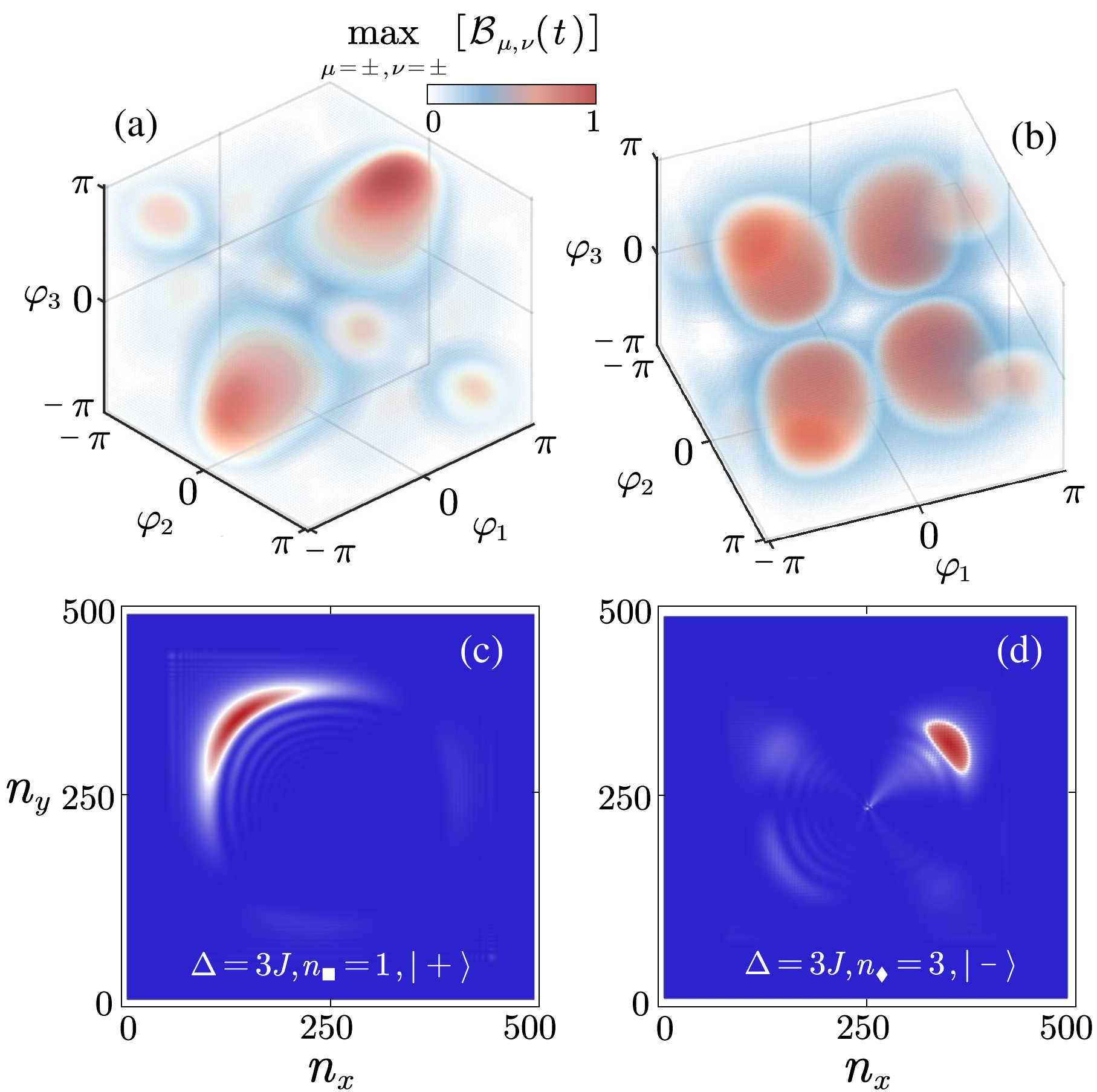}
  \caption{The functions$\max\limits_{\mu=\pm,\nu=\pm}[\mathcal{B}_{\mu,\nu}(t)]$ versus coupling phases $\varphi_{1},\varphi_{2},\varphi_{3}$ for a pair of (a) SGAs and (b) DGAs. Panels (c) and (d) are the bath population $|C_{\boldsymbol{n}}^{\pm}(t)|$ in real space at time $tJ=100$ with $g=0.2J$, utilizing the optimal parameters $\varphi_{p}$ obtained respectively from panels (a) and (b). Other physical parameters are speciﬁed in the legend.}\label{fig3}
  \vspace{-10pt} 
\end{figure}

\section{TUNABLE SUPERRADIANCE AND SUBRADIANCE VIA GA INTERFERENCE}\label{Sec IV}
dynamics of the total GAs plus field is limited in the single-excitation manifold, the state at $t>0$ is given by
\begin{align}\label{eq4}
|\psi(t)\rangle = \left(\sum_{\ell=1,2}C_{\ell}(t)\sigma_{\ell}^{\dagger}+\sum\limits_{\boldsymbol{n}}C_{\boldsymbol{n}}(t)a_{\boldsymbol{n}}^{\dagger}\right)|gg,\{0\}\rangle,
\end{align}
where $C_{\ell}(t)$ and $C_{\boldsymbol{n}}(t)$ are the amplitudes of probability at time $t$ to find an excitation populated in the $\ell$-th atom and in the bosonic mode located at site $\boldsymbol{n} = (n_{x},n_{y})$, respectively.

For emitters initially in a (anti)symmetric superposition $\ket{\pm}$, one can obtain the atomic dynamics by tracing out the reservoir, which yields\,\cite{Cohen_book, sciadv0297}
\begin{align}\label{eq5}
\small
\!\!C_{\pm}(t)=- \frac{1}{2\pi i}\int_{-\infty}^{\infty}dE\,G_{e}^{\pm}(E+i0^{+})e^{-iEt},
\end{align}
where the retarded Green functions take the form of $G_{e}^{\pm}(z)=\left[z-\Delta-\Sigma_{\pm}(z)\right]^{-1}$. In general, the atomic dynamics is fully characterized by the contributions from bound states, unstable poles, and branch-cut-induced detours\,\cite{GonzalezTudela2018,PhysRevLett.125.163602,PhysRevA.103.033511}. The geometry-dependent dynamical contributions for GAs with varying atomic sizes are given in Appendix \ref{B}, enabling several key predictions.

We plot the atomic dynamics governed by Eq.~(\ref{eq5}) for a pair of DGAs [Fig.\,\ref{fig2}(a)] and SGAs [Fig.\,\ref{fig2}(b)], both initialized in state $\ket{-}$ with $\Delta=0$. The former exhibits pronounced subradiance, forming a BIC in the long-time limit under weak coupling. This behavior is explained by a pronounced suppression of decay rates [Fig.\,\ref{fig1}(e)]. The asymptotic BIC occupation probability is given by $\left|1/(1+n^{2}_{\blacklozenge}g^{2}/J^{2})\right|^{2}$, where $n_{\blacklozenge}$ takes only the odd values (see Appendix \ref{C} for more details). For SGAs, the perturbative treatment predicts a divergent decay rate at $\Delta=0$, suggesting a nearly vanishing atomic lifetime. However, as evidenced by Fig.\,\ref{fig2}(b), this behavior is absent, indicating the breakdown of Fermi’s golden rule (FGR). Moreover, we recall that the poles UPII and UPIII coexist within a finite detuning window around the band center [Fig.\,\ref{fig1}(c)]. While these poles share identical decay rates at $\Delta=0$, their real-energy difference induces observable oscillating dynamics.

The excitation revivals of GAs in this 2D bath, especially the emerging dynamical beats in Fig.\,\ref{fig2}(b), motivate a quantitative investigation of non-Markovian memory effects. To quantify it, we adopt the trace distance measure $D(\Lambda_{t}\rho_{1},\Lambda_{t}\rho_{2})$ $=\Vert \Lambda_{t}\rho_{1}\!-\!\Lambda_{t}\rho_{2}\Vert_{1}/2 $ for a pair of evolved states $(\rho_{1},\rho_{2})$ with $\Lambda_{t}$ the dynamical map\,\cite{PhysRevLett.103.210401,PhysRevA.81.062115,PhysRevA.89.012307}. Its positive gradient $\partial_{t}D(\Lambda_{t}\rho_{1},\Lambda_{t}\rho_{2})$ signifies non-Markovian backflow to the emitters. We thus quantify the non-Markovianity $\mathcal{N}(\Lambda)=\mathop{\text{max}}\limits_{\rho_{1}(0),\rho_{2}(0)}\!\!\mathcal{I}\!(\Lambda,\rho_{1},\!\rho_{2})\!$ as the total backflow of information, where the maximization runs over all possible initial state pairs $(\rho_{1},\rho_{2})$, and
\begin{align}\label{eq6}
\mathcal{I}(\Lambda,\rho_{1},\rho_{2})=\int_{\partial_{t}D>0}dt\frac{d D(\Lambda_{t}\rho_{1},\Lambda_{t}\rho_{2})}{dt}
\end{align}
quantifies the distinguishability increase during the evolution.

Figure~\ref{fig2}(c) displays the reduced dynamics of $\mathcal{I}(\Lambda,\rho_{1},\rho_{2})$ for SGAs with $\Delta=0$, where the evolution initiated from $10000$ random quantum state pairs $\ket{\psi_{1}},\ket{\psi_{2}}$ (taking the form of $(\cos\theta\ket{eg}+\sin\theta e^{i\phi}\ket{ge})\otimes \ket{\{0\}}$). We find that the maximum of $\mathcal{N}(\Lambda)$ is attained by the pair of initial states $\ket{+}$ and $\ket{-}$ ($\ket{eg}$ and $\ket{ge}$) during the early (late) stage. Further insight indicates that the product states yield stronger steady non-Markovianity than the entangled states with a slower dynamical response. The similar analyses for DGAs are presented in Appendix \ref{D}.

We can then proceed to study the patterns of emission into the 2D bath, providing a complementary perspective for the dynamical description. The bosonic field density distribution is visualized by solving the real-space photon dynamics via the resolvent technique, which yields\,\cite{PhysRevLett.119.143602,PhysRevA.96.043811,PhysRevLett.129.223601}
\begin{equation}\label{eq7}
\scalebox{0.95}{$\displaystyle
\!\!C_{\boldsymbol{n}}^{\pm}(t)\!=\!\frac{-1}{2\pi i}\int_{-\infty}^{\infty}dE\!\!\int\!\!\!\int \frac{d\boldsymbol{k}}{(2\pi)^{2}}\,G^{\pm}_{\boldsymbol{k}}(E+i0^{+})e^{i(\boldsymbol{k}\cdot\boldsymbol{n}-E)t},
$}
\end{equation}
where the photon Green’s functions are given by $G_{\boldsymbol{k}}^{\alpha}(z)=\sqrt{2\mathcal{P}_{\alpha}(\boldsymbol{k})}gG_{e}^{\alpha}(z)/[z-\omega(\boldsymbol{k})]$.

Figures\,\ref{fig2}(d-g) show the bath population at time $tJ=100$, calculated through numerical integration of Eq.~(\ref{eq7}) for different atomic configurations. Specifically, for a pair of DGAs, we observe directional emission propagating along the principal axes of the 2D square lattice [Fig.\,\ref{fig2}(d)], forming spatially confined photon wave packets. By tuning the atomic geometry to manipulate photonic interference, these wave packets undergo a $\pi/4$ rotation in their propagation direction [Fig.\,\ref{fig2}(e)]. Furthermore, the SGAs initially prepared in $\ket{+}$ exhibit quasi-1D radiation propagating exclusively along one diagonal [Fig.\,\ref{fig2}(f)] when $\Delta=0$. In this scenario, the momentum modes along the lines $k_{x} - k_{y} = \pm \pi$ (mod $2\pi$) are populated primarily (see Appendix \ref{C} for more details). We also capture ripple-like emission [Fig.\,\ref{fig2}(g)], where the energy propagates concentrically from the lattice center. The delicate interference of GAs-emitted photons significantly contributes to these exotic patterns.

\begin{figure}
  \centering
  \includegraphics[width=8.7cm]{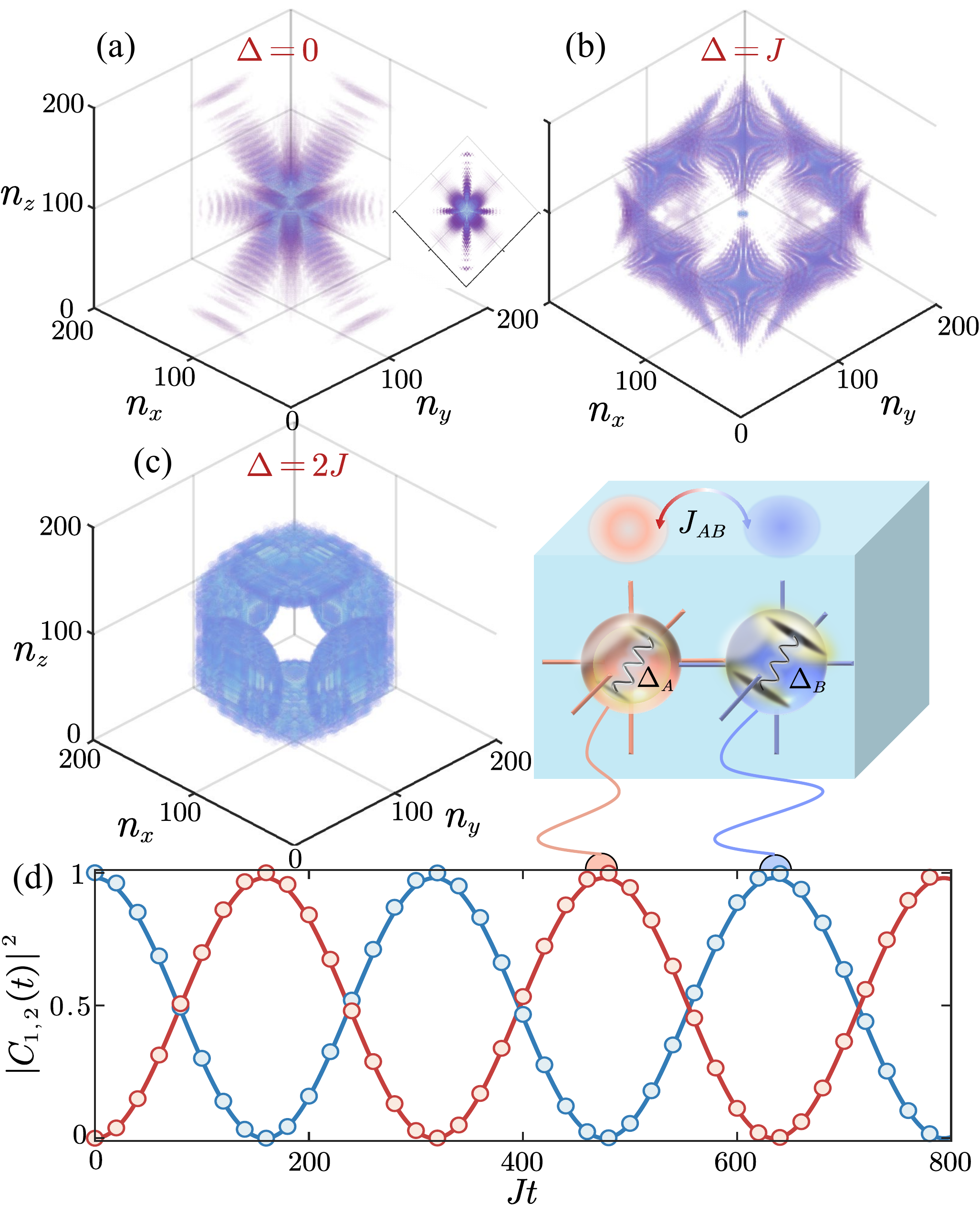}
  \caption{The bath population $|C_{\boldsymbol{n}}^{\pm}(t)|$ in 3D real space at time $tJ=50$ with $g=0.2J$ and $\ket{\psi(0)}=\ket{+}$ for three detuning regimes: $\Delta = 0$ (a), $\Delta = J$ (b), $\Delta = 2J$ (c). An overhead view in the inset of (a) is given. The analytical (markers) and numerical (solid lines) atomic population dynamics $\left|C_{1,2}(t)\right|^{2}$ are shown in panel (c) for the initial $\ket{eg}$ state using parameters $g=0.1J$, $\Delta = 0$, and $n=1$.}\label{fig4}
  \vspace{-10pt} 
\end{figure}

\section{CHIRAL EMISSION}\label{Sec V}
Of fundamental importance, especially in high-dimensional scenario, is the realization of chiral light-matter interfaces \,\cite{PhysRevLett.127.233601,PRXQuantum.6.020101}, which unlocks unprecedented opportunities to probe exotic quantum many-body phenomena, including chiral many-body superradiance\,\cite{PhysRevX.14.011020} and the emergence of correlated multiphoton bound states\,\cite{PhysRevX.10.031011}. Symmetry breaking in GA-waveguide couplings can be achieved  by introducing nontrivial phases $\arg(g_{p})$ at each coupling point\,\cite{PhysRevResearch.4.023198,DuL2023,Qiulaser}, e.g., the couplings $g_{p}$ for SGAs configured in Fig.\,\ref{fig1}(a) take the form of $g(1,e^{i\varphi_{1}},e^{i\varphi_{2}},e^{i\varphi_{3}})$.  In this case, the interference properties are significantly modified, as manifested by the time-dependent fractional power distribution among four directions:
\begin{equation}\label{eq8}
\mathcal{B}_{\mu,\nu}(t) = \sum\limits_{n_{x}^{\mu},n_{y}^{\nu}}\left|C_{\boldsymbol{n}}(t)\right|^{2}/\sum\limits_{\boldsymbol{n}}\left|C_{\boldsymbol{n}}(t)\right|^{2},
\end{equation}
where $n_{x/y}^{\pm}$ denotes the spatial region $n_{x/y}\gtrless\bar{n}_{x/y}$ with $\bar{\boldsymbol{n}}= (\bar{n}_{x},\bar{n}_{y})$ being the central position of the GAs.

Through numerical optimization of the phases $\varphi_{p}$, we maximize the function $\max\limits_{\mu=\pm,\nu=\pm}[\mathcal{B}_{\mu,\nu}(t)]$ at an appropriate instant, e.g., Figs.\,\ref{fig3}(a-b), enabling the identification of collective chiral emission. Specifically, over $90\%$ of the radiated power concentrates within one particular quadrant of the 2D lattice, characterized by the $(n_{x}^{\mu},n_{y}^{\nu})$ sector, as shown in Figs.\,\ref{fig3}(c-d). In contrast, destructive interference of radiated photons leads to significantly diminished power in other quadrants. Chirality in this setting stems from the interference of two phase-sensitive processes: the relative phases between multiple coupling points and the propagation-direction-dependent phase accumulation in this 2D waveguide.
\section{GAS IN 3D ENVIRONMENT}\label{Sec VI}
We now turn to the case where GAs interact with a 3D bath constructed from stacked 2D square lattices, forming a cubic simple lattice with dispersion relation $\omega_{{\rm CS}}(\boldsymbol{k})=-2J[\cos(k_{x})+\cos(k_{y})+\cos(k_{z})]$. It is compelling to study the radiative emission patterns of two GAs: one is centered at the origin and couples to $M=8$ lattice sites at $(s_{x},s_{y},s_{z})$ with $s_{x,y,z}\in\{-n,n\}$, while an identical counterpart is displaced by $(n,n,n)$. We show the photonic excitation distributions at time $tJ = 50$ for $n=1$ and $\ket{\psi(0)}=\ket{+}$ in Fig.\,\ref{fig4}. Specifically, Fig.\,\ref{fig4}(a) presents the case of $\Delta=0$, displaying a highly anisotropic radiation pattern. In Fig.\,\ref{fig4}(b), for $\Delta=J$, the GAs emit radiation forming octahedral patterns with rhombic faces along the principal axes. These rhombic faces transition into squares at $\Delta=2J$, generating a cubic radiation profile [see Fig.\,\ref{fig4}(c)].

\begin{figure}
  \centering
  \includegraphics[width=8.8cm]{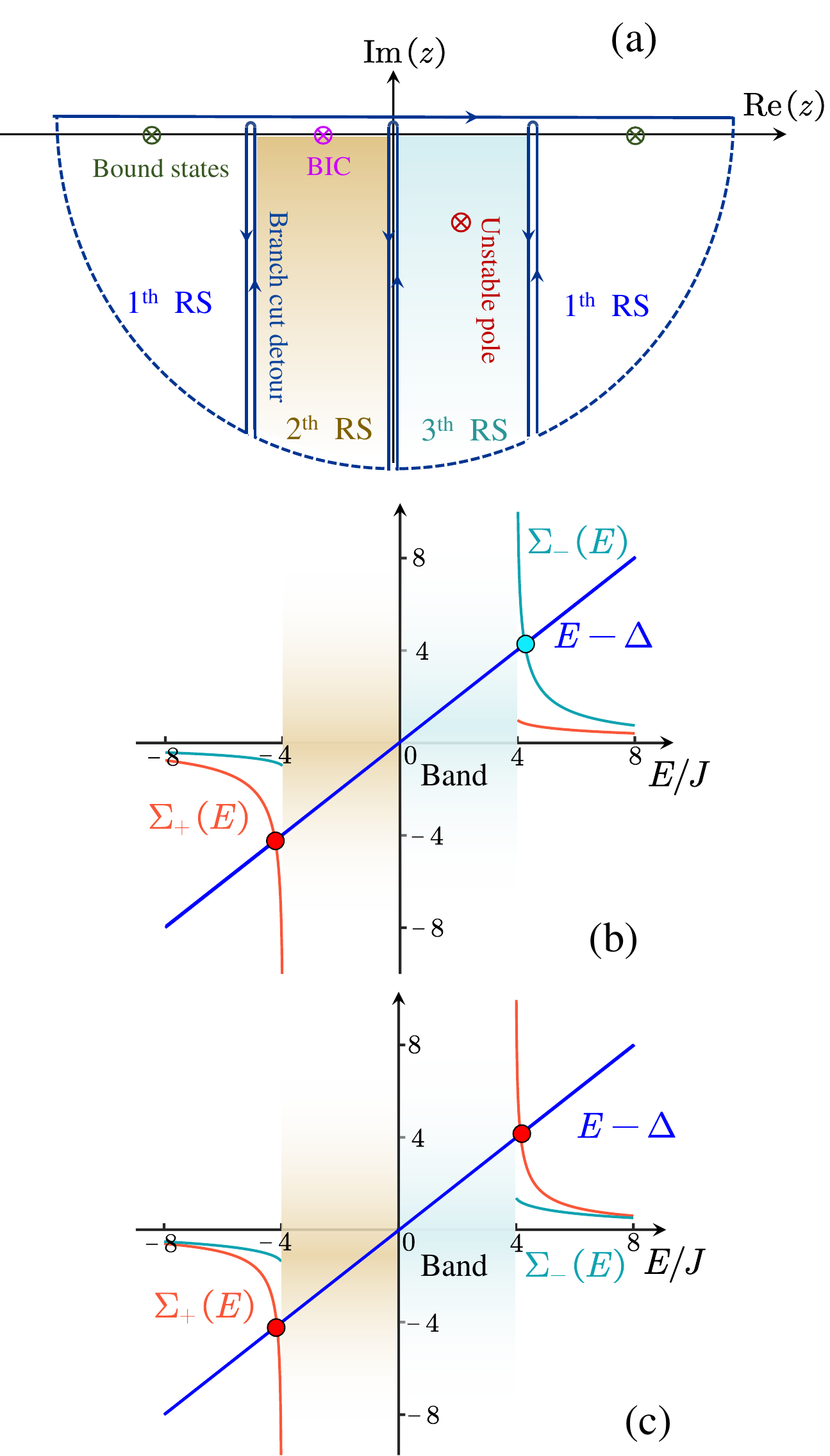}
  \caption{(a) An integration contour (horizontal dark blue line) to calculate Eq.~(\ref{B1}). One needs to close the contour of integration in the lower half of the complex plane (dashed and vertical dark blue line) to evaluate the integration. Here, the information includes bound-state energies, branch cuts detour, unstable poles, and band regions, within the lower half of the complex plane. Near the band edges, the integration contour transitions from the first Riemann sheet ($1^{{\rm st}}$ RS) to the second or third Riemann sheet ($2^{{\rm st}}$ RS and $3^{{\rm st}}$ RS, represented by the brown and cyan regions) when evaluating the integrand in Eq.~(\ref{B1}). Notably, this transition occurs specifically at the band center when moving between the second and third Riemann sheets. Panels (b) and (c) describe the self-energies $\Sigma^{{\rm I}}_{\pm}(E)$ for a pair of DGAs and SGAs, respectively, with $n_{\blacklozenge}=1, n_{\protect\scalebox{0.55}{$\protect\blacksquare$}}=1,\Delta=0$, and $g=J$. Roots of the poles equation $z-\Delta-\Sigma^{{\rm I}}_{\pm}(z)=0$ are obtained from the intersection points between $E-\Delta$ (blue solid line) and $\Sigma^{{\rm I}}_{\pm}(E)$ (red/cyan solid lines).}\label{fig5}
\end{figure}

Among the most remarkable and potentially useful properties of GAs is their capacity for decoherence-free interaction. Inspired by the atomic geometry of DGAs, we consider one emitter connected to $(\pm n,0,0),$ $(0,\pm n,0),(0,0,\pm n)$, with an identical one displaced by $(n,0,0)$. For a pair of GAs spectrally tuned to the band center and prepared in $\ket{eg}$, we observe the emergence of purely coherent dipole-dipole interactions as shown in Fig.\,\ref{fig4}(d) (solid lines). The emergence of undamped oscillations stems from the exact solutions of $z_{\pm}-\Sigma_{\pm}(z_{\pm})=0$, which correspond to stable bound states with energies $z_{\pm}=\pm J_{{\rm AB}}$, where the coupling strength $J_{AB}$ exhibits a marked size dependence ($J_{AB}\approx 0.99 g^{2}$ for $n=1$ versus $J_{AB}\approx 5 g^{2}$ for $n>1$). These non-decaying eigenmodes directly generate the coherent population transfer $\left|C_{1,2}(t)\right|^{2} \approx [1\pm\cos(2J_{ AB}t)]/2$, whose perfect match with numerical simulations is shown in Fig.\,\ref{fig4}(d) (markers).

\section{EXPERIMENTAL IMPLEMENTATION AND SUMMARY}\label{Sec VII}
Regarding experimental implementations, the considered GAs in high-dimensional baths could be realised in dynamical state-dependent optical lattices with ultracold atoms\,\cite{Navarrete2011,Yang2025,Youngshin2025}, where rapid relative displacement between bilayer potentials enables the nonlocal light-matter couplings. Another promising proposal leverages the state-of-the-art superconducting circuits\,\cite{PhysRevA.93.062319,PhysRevB.104.035427,PhysRevA.111.032614}, where qubits are capacitively coupled to a  high-dimensional array of interconnected transmission-line resonators.

To sum up, we have explored the geometry-dependent collective dynamics of GAs bathed in a 2D squared lattice through a non-perturbative description. The nonlocal light-matter interactions lead to a fundamental agreement or discrepancy between the exact descriptions and the predictions from FGR, depending critically on both the atomic configurations and initial states. From a phenomenological standpoint, we also capture several interesting collective radiation patterns generated through GAs' rich interference effects. Remarkably, the photon propagation direction can be precisely engineered to produce chiral emission through controlled phase modulation. Moreover, we establish a 3D extension of GA physics, where numerical simulations reveal two hallmark phenomena, i.e., persistent decoherence-free interactions and dynamically reconfigurable radiation patterns. It  then would be theoretically significant to generalize these results to multi-excitation regime\,\cite{PhysRevLett.125.263601,PhysRevLett.131.033605,PhysRevX.14.011020}, where nonlinear interactions may mediate emergent many-body correlations and enable new phases of quantum light-matter coupling beyond the single-excitation paradigm.

We are grateful to Guoqing Tian for carefully reading the manuscript. This work is supported by the National Science Fund for Distinguished Young Scholars of China (Grant No. 12425502) and the National Key Research and Development Program of China (Grant No. 2021YFA1400700). The computation was completed in the HPC Platform of Huazhong University of Science and Technology.

\begin{figure}
  \centering
  \includegraphics[width=8.8cm]{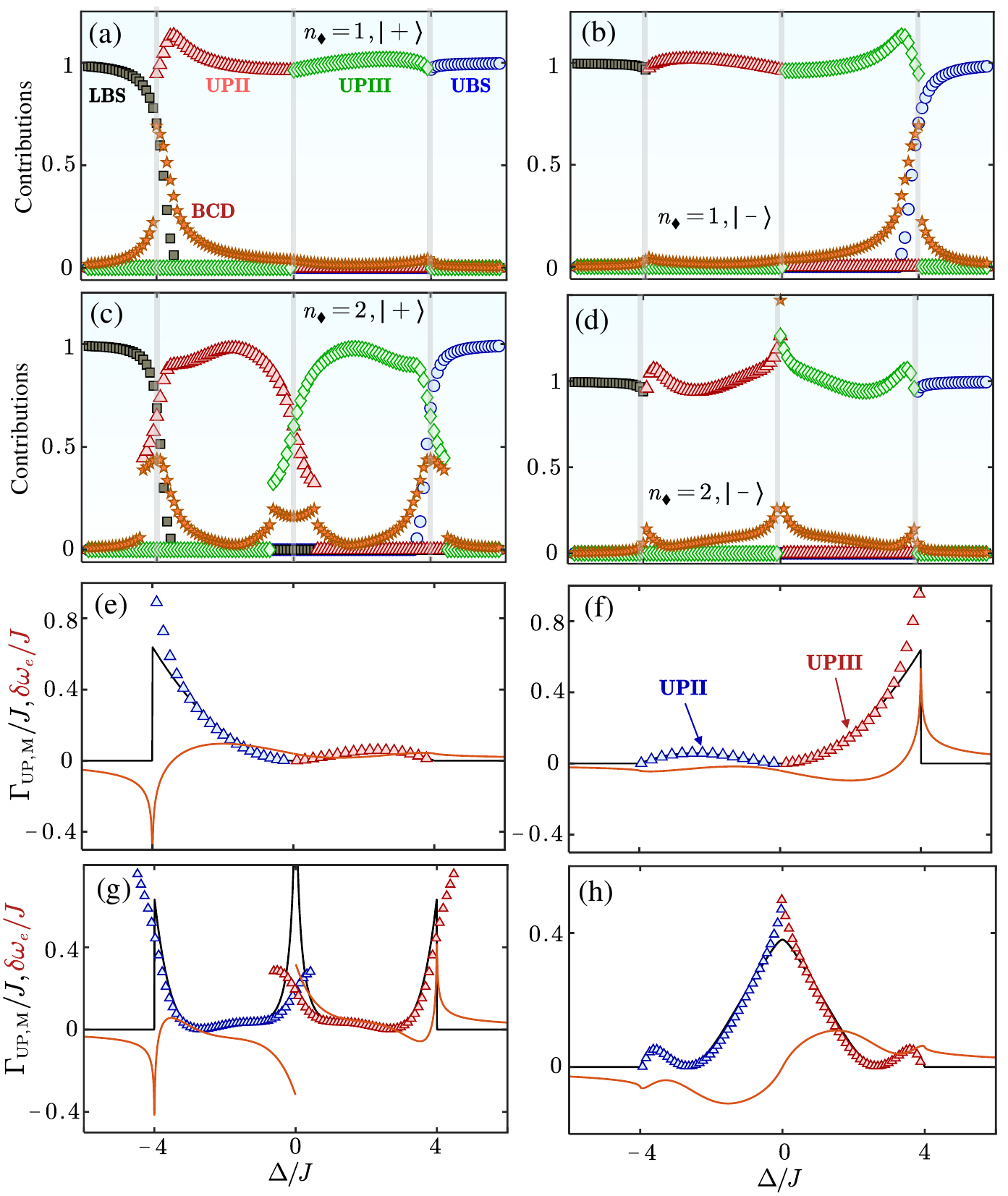}
  \caption{(a) Dynamics contributions of diamond-like GAs for exchange-symmetric [(a) and (c)] and antisymmetric [(b) and (d)] states at time $t = 0$ as a function of detuning $\Delta$. Symbols
  denote: LBS (black square), UBS (blue circles), UPII (red triangles), UPIII (green diamond), BCD (brown stars). The band edges and the band center, i.e., $\Delta=0,\pm 4J$, are marked by grey thick lines. The considered atomic sizes are $n_{\blacklozenge}=1$ for panels (a) and (b) and $n_{\blacklozenge}=2$ for panels (c) and (d), and the light-matter coupling strength reads $g=0.2J$. Panels (e)-(h) are the imaginary parts of the complex poles in the second (blue triangles) and third (green triangles) Riemann sheets versus detuning $\Delta$ compared to Markov prediction (black lines), sharing the same parameters as panels (a)-(d). The collective frequency shifts (red lines) are also plotted for completeness.}\label{fig6}
\end{figure}

\appendix
\renewcommand\appendixname{APPENDIX}
\section{SELF-ENERGIES FOR TWO GIANT QUANTUM EMITTERS} \label{A}
In this section, we give detailed derivation of self-energies $\Sigma_{+}(z)$ and $\Sigma_{-}(z)$, which determine the collective dynamics of a pair of GAs prepared in symmetric and antisymmetric states, respectively. We start by recalling the considered model Hamiltonian $H_{{\rm{tot}}}=H_{A}+H_{B}+H_{{\rm{int}}}$, where $H_{A}=\sum_{\ell=1,2}\Delta\sigma_{\ell}^{\dagger}\sigma_{\ell}^{}$ and $H_{B}=-J\sum_{\langle \boldsymbol{m}, \boldsymbol{n}\rangle}(a_{\boldsymbol{n}}^{\dagger}a_{\boldsymbol{m}}^{}+{\rm{H.c.}})$ are the free Hamiltonians for the GAs and the field, respectively. By making the electric-dipole and rotating wave approximations to neglect the counter-rotating terms, the emitter-field interaction is described by a non-local Hamiltonian $H_{{\rm int}}=\sum_{\ell=1,2}\sum_{p=1}^{M}g_{\ell p}(a_{\boldsymbol{n}_{\ell p}}^{\dagger}\sigma_{\ell}^{}+{\rm H.c.})$, where $\boldsymbol{n}_{\ell p}$ represents the position of the bosonic mode that interacts with the $p\,$-th coupling point of the $\ell$-th atom, at which the coupling strength is given by $g_{\ell p}$. This excitation-conservative Hamiltonian $H_{{\rm{tot}}}$ provides a clear dynamical description.

We proceed by adopting periodic boundary conditions and introducing the operators $a_{\boldsymbol{n}}=\frac{1}{N}\sum_{\boldsymbol{k}} a_{\boldsymbol{k}}e^{i\boldsymbol{k}\cdot\boldsymbol{n}}$, the bath Hamiltonian can be written in momentum space in a diagonal form $H_{B}=\sum_{\boldsymbol{k}}\omega(\boldsymbol{k})a^{\dagger}_{\boldsymbol{k}}a_{\boldsymbol{k}}^{}$ with dispersion relation $\omega(\boldsymbol{k})=-2J[\cos(k_{x})+\cos(k_{y})]$ and discrete wave numbers $k_{x}^{},k_{y}^{}\in \frac{2\pi}{N}(-\frac{N}{2},\cdots,\frac{N}{2}-1)$. The resulting interaction Hamiltonian in the reciprocal space reads
\begin{align}
H_{{\rm{int}}}=\frac{1}{N}\sum_{\ell=1}^{2}\sum_{\boldsymbol{k}}[g_{\ell}(\boldsymbol{k})a_{\boldsymbol{k}}^{\dagger}\sigma_{\ell}+{\rm H.c.}],\label{A1}
\end{align}
where $g_{\ell}(\boldsymbol{k})=\sum_{p=1}^{M}g_{p}e^{i\boldsymbol{k}\cdot\boldsymbol{n_{\ell p}}}$ is the $\boldsymbol{k}$-dependent coupling function. To elucidate the collective dynamics, we introduce the orthogonal mode operators $\sigma_{\pm}\equiv(\sigma_{1}\pm\sigma_{2})/\sqrt{2}$, which provide a more transparent description of the system's behavior. The interaction Hamiltonian (\ref{A1}) in these new basis states takes the form
\begin{align}
H_{{\rm{int}}}=&\frac{1}{\sqrt{2}N}\sum_{\boldsymbol{k}}\{g_{+}(\boldsymbol{k})a_{\boldsymbol{k}}^{\dagger}\sigma_{+}+g_{-}(\boldsymbol{k})a_{\boldsymbol{k}}^{\dagger}\sigma_{-}+{\rm H.c.}\}, \nonumber\\
=&\frac{1}{\sqrt{2}N}\sum_{\boldsymbol{k}>0}\sum_{\alpha=\pm}(\sqrt{N_{\alpha}(\boldsymbol{k})}a_{\boldsymbol{k},\alpha}^{\dagger}\sigma^{}_{\alpha}+{\rm H.c.}),\label{A2}
\end{align}
where the notation $g_{\pm}(\boldsymbol{k})\!\equiv\! g_{1}(\boldsymbol{k})\!\pm\! g_{2}(\boldsymbol{k})$ is introduced, and the bosonic modes $a_{\boldsymbol{k},\pm}$ are defined as $a_{\boldsymbol{k},\pm}\equiv\frac{1}{\sqrt{N_{\pm}(\boldsymbol{k})}}\sum\limits_{\alpha,\alpha'=\pm}\left[ g^{*}_{1}(\alpha k_{x},\alpha' k_{y})\pm g^{*}_{2}( \alpha k_{x}, \alpha' k_{y})\right]a_{\alpha k_{x},\alpha' k_{y}}$, satisfying bosonic commutation relation $[a_{\boldsymbol{k},\alpha},a_{\boldsymbol{k},\alpha'}^{\dagger}]=\delta_{\alpha,\alpha'}\delta(\boldsymbol{k}-\boldsymbol{k}')$. Here, the normalized constants $N_{\pm}(\boldsymbol{k})$ read
\begin{align}
\!\!\!\!N_{\pm}(\boldsymbol{k})\!=&4g^{2}\Big\{2M\!+\!2\sum_{p\neq p'}\big[\cos(\boldsymbol{k}\!\cdot\!\triangle\boldsymbol{n}_{pp'}^{11})+\cos(\boldsymbol{k}^{*}\!\cdot\!\triangle\boldsymbol{n}_{pp'}^{11})\big]\!\nonumber\\
&\! \pm\!\!  \sum_{p,p'}\cos(\boldsymbol{k}\!\cdot\!\triangle\boldsymbol{n}_{pp'}^{12})\!\pm\!\sum_{p,p'}\cos(\boldsymbol{k}^{*}\!\cdot\!\triangle\boldsymbol{n}_{pp'}^{12})\Big\}\! ,\label{A3}
\end{align}
where $\Delta\boldsymbol{n}_{pp'}^{\ell \ell^{\prime}}$ is the relative coordinate from the $p\,$-th point of atom $\ell$ to the $p'$-th point of atom $\ell'$. The transformed wavevector $\boldsymbol{k}^{*} \equiv (-k_x, k_y)$ is defined by inverting the sign of the $x$-component of momentum $\boldsymbol{k} = (k_x, k_y)$. Note that the interaction Hamiltonian (\ref{A2}) can be rewritten in the form:
\begin{align}
H_{{\rm int}}=\frac{\sqrt{2}g}{N}\sum_{\boldsymbol{k}\in\boldsymbol{k}^{{\rm I}}}\sum_{\alpha=\pm^{}}[\sqrt{\mathcal{P}_{\alpha}(\boldsymbol{k})}a_{\boldsymbol{k},\alpha}^{\dagger}\sigma^{}_{\alpha}+{\rm H.c.}],\label{A4}
\end{align}
which is exactly the Eq.~(\ref{eq1}) in the main text with $\mathcal{P}_{\alpha}(\boldsymbol{k})=\mathcal{N}_{\alpha}(\boldsymbol{k})/4g^{2}$. Obviously, the bath modes $a_{\boldsymbol{k},\pm}$ are mutually orthogonal, leading to decoupled dynamics for the collective states $\ket{\pm}=\sigma_{\pm}^{\dagger}\ket{gg}$.  This decoupling allows the dynamics of states $\ket{\pm}$ to be analyzed independently.

\begin{figure}
  \centering
  \includegraphics[width=8.8cm]{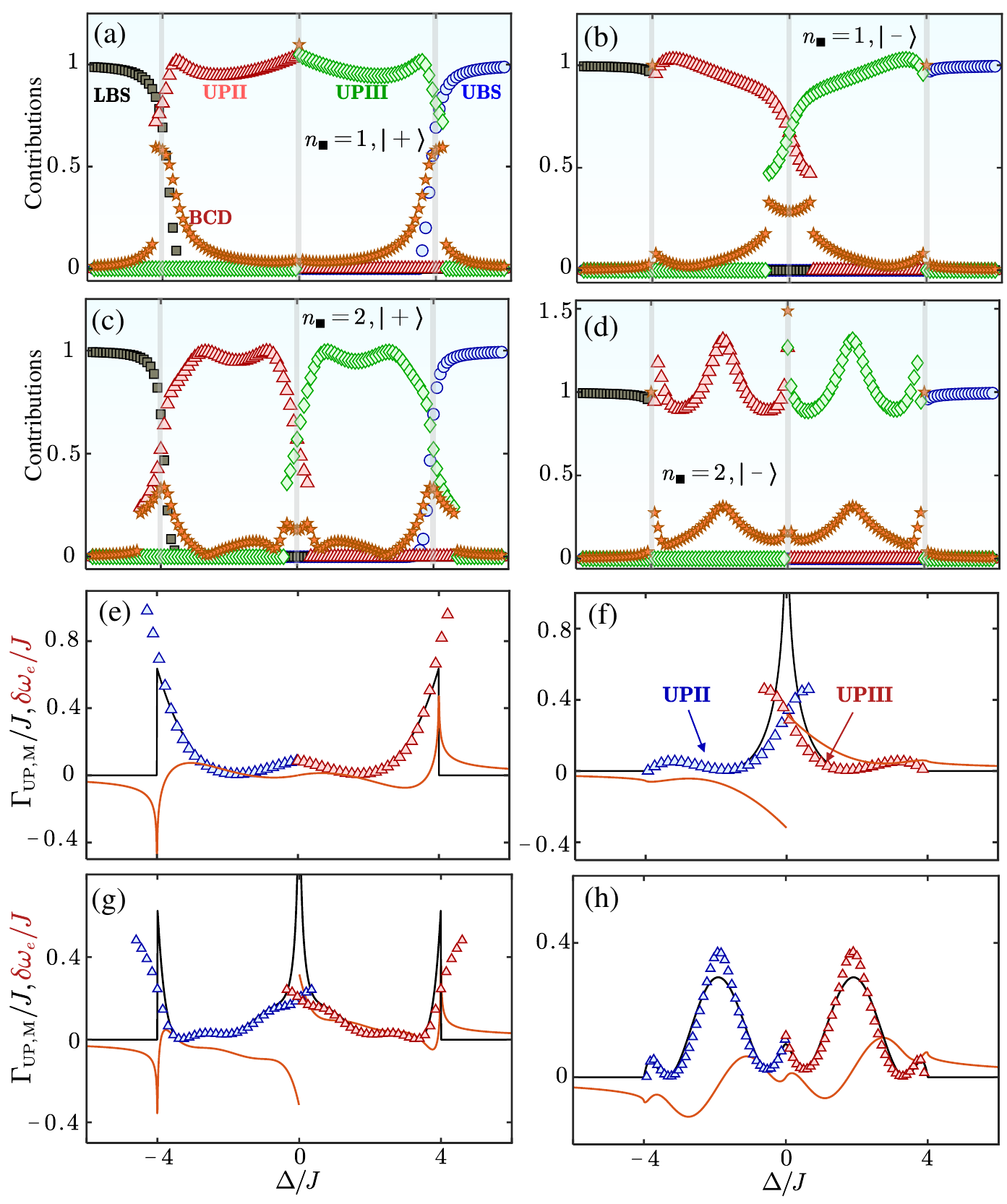}
  \caption{(a) Dynamics contributions of square-like GAs for exchange-symmetric [(a) and (c)] and antisymmetric [(b) and (d)] states at time $t = 0$ as a function of detuning $\Delta$. Symbols
  denote: LBS (black square), UBS (blue circles), UPII (red triangles), UPIII (green diamond), BCD (brown stars). The band edges and the band center, i.e., $\Delta=0,\pm 4J$, are marked by grey thick lines. The considered atomic sizes are $n_{\protect\scalebox{0.55}{$\protect\blacksquare$}}=1$ for panels (a) and (b) and $n_{\protect\scalebox{0.55}{$\protect\blacksquare$}}=2$ for panels (c) and (d), and the light-matter coupling strength reads $g=0.2J$. Panels (e)-(h) are the imaginary parts of the complex poles in the second (blue triangles) and third (green triangles) Riemann sheets versus detuning $\Delta$ compared to Markov prediction (black lines), sharing the same parameters as panels (a)-(d). The collective frequency shifts (red lines) are also plotted for completeness.}\label{fig7}
\end{figure}

We are now ready to decompose the complete Hilbert space into the subspace of interest $\mathcal{P}$ and its complementary subspace $\mathcal{Q}=1-\mathcal{P}$. The resolvent $G(z) = (z-H_{{\rm{tot}}})^{-1}$ can be projected onto\,\cite{SMCohen_book, SMsciadv0297}
\begin{align}
\mathcal{P}G(z)\mathcal{P}&=\frac{\mathcal{P}}{z-E_{\mathcal{P}}-\Sigma_{\mathcal{P}}(z)},\label{A5}\\
\!\!\!\mathcal{Q}G(z)\mathcal{P}&=\frac{\mathcal{Q}}{z\!-\!E_{\mathcal{Q}}\!-\!\Sigma_{\mathcal{Q}}(z)}H_{{\rm int}}\frac{\mathcal{P}}{z\!-\!E_{\mathcal{P}}\!-\!\Sigma_{\mathcal{P}}(z)},\label{A6}
\end{align}
where the notations $E_{X}\equiv X(H_{A}+H_{B})X$ and $\Sigma_{X}(z)=X\boldsymbol{\Sigma}(z)X$ have been introduced for simplicity. Here, $\boldsymbol{\Sigma}(z)$ is the level-shift operator and is given by
\begin{align}\label{A7}
\boldsymbol{\Sigma}(z) = H_{{\rm int}}+H_{{\rm int}}\frac{\mathcal{Q}}{z-E_{\mathcal{Q}}-\mathcal{Q}H_{{\rm int}}\mathcal{Q}}H_{{\rm int}}.
\end{align}

It is now possible to obtain the self-energy for a symmetric (antisymmetric) state when we choose $\mathcal{P}=\ket{+}\bra{+}\ (\ket{-}\bra{-})$, which takes the simple form of $\Sigma_{\pm}(z)\equiv\bra{\pm}\boldsymbol{\Sigma}(z)\ket{\pm}=\Sigma_{e}(z)\pm\Sigma_{12}(z)$, with
\begin{align}
\Sigma_{e}(z)&=\frac{g^{2}}{N^{2}}\sum_{\boldsymbol{k}}\frac{M+2\sum_{p\neq p'}\cos(\boldsymbol{k}\cdot\Delta\boldsymbol{n}_{pp'}^{11})}{z-\omega(\boldsymbol{k})},\label{A8}\\
\Sigma_{12}(z)&=\frac{g^{2}}{N^{2}}\sum_{p,p'}\sum_{\boldsymbol{k}}\frac{\cos(\boldsymbol{k}\cdot\Delta\boldsymbol{n}_{pp'}^{12})}{z-\omega(\boldsymbol{k})}.\label{A9}
\end{align}
 The derivations presented above rely on the fact that the two GAs share the identical spatial geometries. For a given spatial configuration of giant emitters, the self-energies in Eqs.~(\ref{A8}) and (\ref{A9}) become analytically tractable when all quantities $\Sigma(z;[m,n])$, defined by
\begin{align}
\Sigma(z;[m,n]) = \frac{g^{2}}{(2\pi)^{2}}\iint d\boldsymbol{k}\frac{\cos(k_{x}m+k_{y}n)}{z+2J(\cos k_{x}+\cos k_{y})},\label{A10}
\end{align}
are pre-computed. Given the following basic self-energy components
\begin{align}
\Sigma(z;[0,0]) =& \frac{2g^{2}}{\pi z}K[m(z)], \nonumber\\
\Sigma(z;[1,1]) =& \frac{2g^{2}}{\pi z}\left\{ (\frac{2}{m(z)}-1)K[m(z)]-\frac{2}{m(z)}E[m(z)]\right\},\nonumber\\
\Sigma(z;[1,0]) =& \frac{g^{2}}{4J}-\frac{g^{2}}{2\pi J}K[m(z)],\label{A11}
\end{align}
where $m(z)\equiv (4J/z)^{2}$, the expressions of self-energies $\Sigma(z;[m,n])$ for arbitrary $[m,n]$ are available by performing the following recursive relations\,\cite{Morita,Guttmann_2010}:
\begin{widetext}
\begin{align}
\Sigma(z;[n+1,0]) =& -\frac{1}{2J}[2z\Sigma(z;[n,0])+2J\Sigma(z;[n-1,0]) +4J\Sigma(z;[n,1])],\nonumber\\
\Sigma(z;[n+1,n]) =& -\frac{1}{4J}[2z\Sigma(z;[n,n])+4J\Sigma(z;[n,n-1])],\nonumber\\
\Sigma(z;[n+1,n+1]) =& \frac{4n}{2n+1}[\frac{2}{m(z)}-1]\Sigma(z;[n,n])-\frac{2n-1}{2n+1}\!\Sigma(z;[n-1,n-1]).\label{A12}
\end{align}
\end{widetext}
The special functions $K(\bullet)$ and $E(\bullet)$ in Eq. (\ref{A11}) represent the complete elliptical integral of the first and second kind, respectively. For illustration purpose, the analytical expressions for the self-energies of a pair of DGAs with $n_{\blacklozenge}=1$ can be derived from Eqs. (\ref{A11}) and  (\ref{A12}) as
\begin{widetext}
\begin{align}\label{A13}
\Sigma_{\pm}(z) = &\,4\Sigma(z;[0,0])+4\Sigma(z;[2n_{\blacklozenge},0])+8\Sigma(z;[n_{\blacklozenge},n_{\blacklozenge}])\pm\left[9\Sigma(z;[n_{\blacklozenge},0])+6\Sigma(z;[2n_{\blacklozenge},n_{\blacklozenge}])+\Sigma(z;[3n_{\blacklozenge},0])\right],\nonumber\\
= &\,\frac{g^{2}}{4J}\left(\frac{4z}{J}\mp\frac{z^{2}}{J^{2}}\right)\left(\frac{2}{\pi }K\left[\left(\frac{4J}{z}\right)^{2}\right]-1\right)\pm\frac{g^{2}}{J}.
\end{align}
Similarly, the self-energies for a pair of SGAs with $n_{\protect\scalebox{0.55}{$\protect\blacksquare$}}=1$ read
\begin{align}\label{A14}
\Sigma_{\pm}(z) = &\,4\Sigma(z;[0,0])+4\Sigma(z;[2n_{\protect\scalebox{0.55}{$\protect\blacksquare$}},2n_{\protect\scalebox{0.55}{$\protect\blacksquare$}}])+8\Sigma(z;[2n_{\protect\scalebox{0.55}{$\protect\blacksquare$}},0])
\pm\left[9\Sigma(z;[n_{\protect\scalebox{0.55}{$\protect\blacksquare$}},n_{\protect\scalebox{0.55}{$\protect\blacksquare$}}])+6\Sigma(z;[3n_{\protect\scalebox{0.55}{$\protect\blacksquare$}},n_{\protect\scalebox{0.55}{$\protect\blacksquare$}}])+\Sigma(z;[3n_{\protect\scalebox{0.55}{$\protect\blacksquare$}},3n_{\protect\scalebox{0.55}{$\protect\blacksquare$}}])\right],\nonumber\\
= &-\frac{16}{3}\Sigma(z;[0,0])-\frac{8z}{J}\Sigma(z;[1,0])+[\frac{32}{3}\frac{1}{m(z)}-20]\Sigma(z;[1,1])\pm\nonumber\\
&\!\!\left\{(\frac{128}{15}\!-\!\frac{16}{15m(z)})\Sigma(z;[0,0])\!+\!\frac{12z}{J}\Sigma(z;[1,0])\!+\!\left(\frac{3z^{2}}{J^{2}}\!+\!12\!+\!\frac{8}{15}(\frac{2}{m(z)}\!-\!1)(\frac{8}{m(z)}\!-\!19)\right)\Sigma(z;[1,1])\right\}.
\end{align}
\end{widetext}
In particular, the real and imaginary parts of the self-energies $\Sigma_{\pm}(\Delta+i0^{+})$ above the real axis correspond to the collective frequency shifts and the effective relaxation rates, respectively, of the entangled states $\ket{\pm}$.
\section{DYNAMICS OF TWO GIANT QUANTUM EMITTERS IN THE NON-MARKOVIAN REGIME}\label{B}
In this section, we present detailed descriptions of atomic and photonic dynamics based on the resolvent formalism and the relevant self-energies given in Sec.\ref{A}. Notably, the projections in Eqs.\,(\ref{A5}) and (\ref{A6}) are essential in simulating the dynamic of the atomic and photonic parts, respectively, from which the atomic probability amplitudes $C_{e}^{\pm}(t)$ and the photonic amplitudes $C^{\pm}_{\boldsymbol{n}}(t)$ in the real space are given by
\begin{align}
C_{e}^{\pm}(t)&=-\frac{1}{2\pi i}\int_{-\infty}^{\infty}dE\,G_{e}^{\pm}(E+i0^{+})e^{-iEt},\label{B1}\\
C_{\boldsymbol{n}}^{\pm}(t)&=-\frac{1}{2\pi i}\int_{-\infty}^{\infty}dE \iint \frac{d\boldsymbol{k}}{(2\pi)^{2}}\,G^{\pm}_{\boldsymbol{k}}(E+i0^{+})e^{i(\boldsymbol{k}\cdot\boldsymbol{n}-E)t},\label{B2}
\end{align}
which are exactly the Eqs.~(\ref{eq5}) and (\ref{eq7}) in the main text, respectively. The retarded Green functions in Eqs.\,(\ref{B1}) and (\ref{B2}) have the form of
\begin{align}
G_{e}^{\pm}(z)=\frac{1}{z-\Delta-\Sigma_{\pm}(z)},\label{B3}\\
G_{\boldsymbol{k}}^{\pm}(z)=\frac{g\sqrt{2\mathcal{P}_{\alpha}(\boldsymbol{k})}}{[z-\omega(\boldsymbol{k})][z-\Delta-\Sigma_{\pm}(z)]}.\label{B4}
\end{align}
To further obtain the atom decay and photon emission dynamics, i.e., calculating the integrals in Eqs.~(\ref{B1}) and (\ref{B2}), we employ residue integration by closing the contour in the lower half of the complex plane, as illustrated in Fig.~\ref{fig5}(a). Now, let us focus on the integral in Eq.~(\ref{B1}) and its integrand. Since the integrand has branch cuts in the real axis along the region (i.e., $z\in [-4J, 4J]$), which corresponds to the continuous spectrum of $H_{B}$.  According to the residue theorem, the integral can be decomposed into contributions from poles and branch cuts. While pole contributions are readily obtained by calculating residues, the branch cut treatment requires more sophisticated approaches. We present two principal methods for handling the branch cuts. The first involves direct integration along a tightly enclosing contour around the branch cuts. The second, and our preferred approach, employs analytic continuation by detouring around band edges into other Riemann sheets of the integrand, as illustrated by the branch cut detours in Fig.~\ref{fig5}(a). This method reveals that the integrand may possess unstable poles in other Riemann sheets, thereby decomposing the branch cut contribution into parts from these poles and branch cut detours.  The analytical expressions for the self-energy, including those in Eqs.~(\ref{A13}) and (\ref{A14}), correspond to the integrand in the first Riemann sheet, denoted $\Sigma^{{\rm I}}_{\pm}(z)$, thereby defining the associated Green's functions $G^{\pm,{\rm I}}_{e}(z)$. They can be analytically continued to the second and third Riemann sheets ($\Sigma^{{\rm II}/{\rm III}}_{\pm}(z)$ and correspondingly $G^{\pm,{\rm II}/{\rm III}}_{e}(z)$, depicted by the brown/cyan areas) through the transformations: $K(m)\rightarrow K(m) \pm 2iK(1-m)$ and $E(m)\rightarrow E(m) \pm  2i[K(1-m)-E(1-m)]$ in the self-energy expressions\,\cite{SMGonzalezTudela2018}. This specific linear combination of elliptic integrals is carefully selected to ensure the self-energy remains continuous across the entire integration contour. The above analytic continuation provides a powerful tool for examining the system's behavior across different Riemann sheets.

We first perform an elaborate characterization of the atom-photon bound states, which are crucial for understanding the long-time dynamical behavior.  As shown in Figs.~\ref{fig5}(b) and (c), for $\Delta=0$, we plot the functions $E-\Delta$ and self-energies $\Sigma^{{\rm I}}_{\pm}(E)$ that are defined in the first Riemann sheet as a function of real energy $E$, comparing results for a pair of DGAs ($n_{\blacklozenge}=1$) and SGAs ($n_{\protect\scalebox{0.55}{$\protect\blacksquare$}}=1$), respectively. The intersection points between the two functions correspond to the emergence of bound states, whose number is strongly dependent on both the coupling strength $g$ and detuning $\Delta$. For a pair of DGAs initially prepared in states $\ket{\pm}$, the system guarantees at least one bound state energy and exhibits two intersection points when $\Delta\lessgtr \pm 4J-\Sigma^{{\rm I}}_{\pm}(\pm 4J)$. In contrast, for a pair of SGAs initially prepared in states $\ket{+}$, the system always possesses a pair of bound states, while no bound states exist for the exchange-antisymmetric state $\ket{-}$ when detuning $\Delta\in(-4J-\Sigma^{{\rm I}}_{-}(-4J),4J-\Sigma^{{\rm I}}_{-}(4J))$. Beyond the number of bound states, their residues play a more fundamental role in governing atomic dynamics, as we will demonstrate below.

In general, the atomic population dynamics is fully characterized by bound states (BSs), unstable poles (UPs), and branch-cut-induced detours (BCDs) contributions:
\begin{align}\label{B5}
C_{\pm}\left( t \right) =\sum_{\beta =\mathrm{BS},\mathrm{UP}}{R^{\pm}_{\beta}}e^{-iz^{\pm}_{\beta}t}+C^{\pm}_{\mathrm{BCD}}\left( t \right),
\end{align}
where $R^{\pm}_{\beta}$ represents the overlap between the initial wave function $\ket{\pm}$ and either the BSs or UPs, obtained via standard contour-integration techniques, and the poles $z^{\pm}_{{\rm BS}}$ and $z^{\pm}_{{\rm UPII/UPIII}}$ are determined by solving the pole equations $z^{\pm}_{{\rm BS}}-\Delta-\Sigma^{{\rm I}}_{\pm}(z^{\pm}_{{\rm BS}})=0$ and $z^{\pm}_{{\rm UPII/UPIII}}-\Delta-\Sigma^{{\rm II/III}}_{\pm}(z^{\pm}_{{\rm UPII/UPIII}})=0$, respectively. The explicit forms of Residue $R^{\pm}_{\beta}$ and branch-cut-induced detours $C^{\pm}_{\mathrm{BCD}}(t)$ are given by\,\cite{SMPhysRevLett.119.143602, SMPhysRevA.96.043811}:
\begin{align}
R^{\pm}_{\beta} =& \left|\frac{1}{1-\partial_{z}\Sigma_{\pm}(z)}\right|_{z=z^{\pm}_{\beta}},\label{B6}
\end{align}
\begin{widetext}
\begin{align}
C^{\pm}_{\mathrm{BCD}}(t) =& \frac{1}{2\pi}\sum\limits_{j=1}^{3}\int_{-\infty}^{0}dy [G_{e}^{\pm,\mathrm{R_{j}}}(x_{j}+iy)-G_{e}^{\pm,\mathrm{L_{j}}}(x_{j}+iy)]e^{-i(x_{j}+iy)t},\label{B7}
\end{align}
\end{widetext}
where the notations $\boldsymbol{{\rm R}}=(\mathrm{I,III,II}), \boldsymbol{\rm{L}}=(\mathrm{III,II,I}), \boldsymbol{x}=(4J,0,-4J)$ have been introduced for simplicity. Here, the labels ${\rm UPII}$ and ${\rm UPIII}$ indicate unstable poles in the second and third Riemann sheets.

For sufficiently weak coupling $g$ in the Markovian limit, the self-energies $\Sigma_{\alpha}(E+i0^{+})$ in Eqs. (\ref{A13}) and (\ref{A14}) can be approximated by evaluating it at the bare energy $\Delta$:
\begin{align}\label{B8}
\small
\!\!\!\!\Sigma_{\alpha}(E+i0^{+})\approx\Sigma_{\alpha}(\Delta+i0^{+})=\delta\omega^{\alpha}_{e}(\Delta)\!-\!i\frac{\Gamma^{\alpha}_{M}(\Delta)}{2},
\end{align}
where $\delta\omega^{\alpha}_{e}(\Delta)$ and $\Gamma^{\alpha}_{M}(\Delta)$ denote the Markovian collective frequency shift and decay rate, respectively. Consequently, the integration of dynamics in Eq.(\ref{B3}) can be performed straightforwardly, yielding immediately an exponential decay $e^{-\Gamma^{\alpha}_{M}(\Delta)t}$ for atomic population. However, we re-emphasize that such perturbative treatment breaks down when the detuning $\Delta$ nears the band edges or the middle of the band, necessitating exact evaluation of the Fourier integrals in Eqs.(\ref{B1})-(\ref{B2}) beyond perturbation theory.

Figure\,\ref{fig6} presents the initial-time $(t=0)$ absolute contributions in Eq.\ref{B5} alongside a comparative analysis of Markovian and non-Markovian decay rates for a pair of DGAs under size variation, from which we derive the fundamental physics of GA dynamics. The principal findings can be summarized as follows:

(i) In the lower band edge, the symmetric lower bound state (LBS) contribution survives for the whole parameter range for both $n_{\blacklozenge}=1$ and $n_{\blacklozenge}=2$. The existence of a bound state below the lower band edge is guaranteed by the function $F(E)\equiv  E-\Delta-\Sigma^{{\rm I}}_{+}(E+i0^{+})$ satisfying $F(E=-4J)>0$ (due to $\lim\limits_{E\rightarrow -4J}\Sigma^{{\rm I}}_{+}(E)=-\infty$) and $F(E=-\infty)<0$, as shown in Fig.\,\ref{fig5}(b).

(ii) In the upper band edge, the contribution of the symmetric upper bound state (UBS) vanishes at a critical detuning $\Delta_{c} = 4J-\Sigma^{{\rm I}}_{+}(4J)$ for $n_{\blacklozenge}=1$, but persists across the entire parameter range for $n_{\blacklozenge}=2$, a behavior fully explained by the bound-state quantitative analysis in Fig.\,\ref{fig5}. The anti-symmetric UBS contribution exhibits an inverse behavior relative to the symmetric case, as ensured by the origin-symmetry between the self-energies $\Sigma^{{\rm I}}_{+}(E)$ and $\Sigma^{{\rm I}}_{-}(E)$. In other words, the merging of the BS into the continuum alternates between the symmetric and antisymmetric components for a fixed $n_{\blacklozenge}$.

(iii) In the middle of the band, the UP contribution dominates in both the symmetric and antisymmetric components for $n_{\blacklozenge}=1$, whereas a non-negligible branch-cut contribution appears for $n_{\blacklozenge}=2$. The contributions of unstable poles ${\rm UPII}$ and ${\rm UPIII}$ are asymmetric and symmetric with respect to the band center for $n_{\blacklozenge}=1$ and $n_{\blacklozenge}=2$, respectively. Moreover, both UPII and UPIII contributions coexist within a finite energy window around the band center for state $\ket{+}$ with $n_{\blacklozenge}=2$, with these unstable poles persisting simultaneously in the second and third Riemann sheets. The coexistence of these poles originates from the finite value of $\delta\omega^{+}_{e}(\Delta)$ around $\Delta \gtrless 0$, which constrains the UP energies to remain within the second and third Riemann sheets.

(iv) Both Markovian and non-Markovian collective decay rates display size-dependent symmetries: asymmetric (for $n_{\blacklozenge}=1$) versus symmetric (for $n_{\blacklozenge}=2$) about the band center. Specifically, for $n_{\blacklozenge}=1$, the non-Markovian decay rates of states $\ket{\pm}$ show enhanced agreement with Markovian predictions near opposite band edges (upper edge for $\ket{+}$, lower edge for $\ket{-}$). For $n_{\blacklozenge}=2$, however, this agreement optimizes at central Riemann sheet regions (for $\ket{+}$) versus band edges (for $\ket{-}$). In addition, the decay rates exhibit size-dependent modulation near $\Delta=0$, which exhibit suppression (enhancement) behavior corresponding to subradiance (superradiance) for $n_{\blacklozenge}=1$ ($n_{\blacklozenge}=2$). Remarkably, a bound state in the continuum (BIC) emerges at the band center for the subradiant case, exhibiting exceptional longevity.

We are now in a position to study the dynamical features of SGAs, following above mentioned procedures. As shown in Fig.\,\ref{fig7}, we plot the contributions to the dynamics at $t=0$ as well as the comparisons between Markovian and non-Markovian decay rates for a pair of SGAs by considering different atomic size and initial states, from which we obtain several basic observations:

(i) In the band edges, the symmetric LBS and UBS contributions survive for the whole parameter range for both $n_{\protect\scalebox{0.55}{$\protect\blacksquare$}}=1$ and $n_{\protect\scalebox{0.55}{$\protect\blacksquare$}}=2$. The existence of bound states outside the band edges is guaranteed by the conditions $F(E=\pm 4J)\lessgtr 0$  and $F(E=\pm\infty)\gtrless0$, as shown in Fig.\,\ref{fig5}(c). The anti-symmetric LBS and UBS contributions vanish at critical detunings $\Delta_{c} = -4J-\Sigma^{{\rm I}}_{-}(-4J)$ and $\Delta_{c} = 4J-\Sigma^{{\rm I}}_{-}(4J)$, respectively.

(ii) In the middle of the band, both the UP and BCD contributions are essential in governing the dynamics of symmetric and antisymmetric states with arbitrary atomic size $n_{\protect\scalebox{0.55}{$\protect\blacksquare$}}$. The contributions of unstable poles ${\rm UPII}$ and ${\rm UPIII}$ are always symmetric with respect to the band center in this situation. Meanwhile, the antisymmetric (symmetric) UPII and UPIII contributions coexist within a finite energy window around the band center for $n_{\protect\scalebox{0.55}{$\protect\blacksquare$}}=1$ ($n_{\protect\scalebox{0.55}{$\protect\blacksquare$}}=2$), i.e., these unstable poles persisting simultaneously in the second and third Riemann sheets. Such a coexistence behavior only occurs in the band center for the symmetric (antisymmetric) case with $n_{\protect\scalebox{0.55}{$\protect\blacksquare$}}=1$ ($n_{\protect\scalebox{0.55}{$\protect\blacksquare$}}=2$).

(iii) Both Markovian and non-Markovian collective decay rates are symmetric about the band center. In particular, for $n_{\protect\scalebox{0.55}{$\protect\blacksquare$}}=1$, the non-Markovian decay rates of state $\ket{+}/\ket{-}$ show enhanced agreement with Markovian predictions near the band center/edges, whereas for $n_{\protect\scalebox{0.55}{$\protect\blacksquare$}}=2$, this agreement optimizes in center/off-center regions of the second and third Riemann sheets for $\ket{+}/\ket{-}$. Notably, the collective decay rates for state $\ket{+}$ extend beyond the band edges $(|E|>4J)$, while for state $\ket{-}$, they remain strictly confined within the band region $(|E|\le 4J)$.
\section{BOUND STATES IN THE CONTINUUM AND  EMISSION PATTERNS}\label{C}
In this section, we determine the parameter regimes supporting bound states in the continuum, and study the exotic lattice dynamics both in real and momentum space. We recall that a pair of DGAs coupled to a 2D bath can support BICs, even when the detuning energy $\Delta$ is precisely at the band center. In the following analysis, we systematically determine (i) the asymptotic behavior in the long-time limit of these special bound states and (ii) their existence conditions. For a pair of DGAs, their symmetric and anti-symmetric self-energies $\Sigma_{\pm}(z)$ admit the following integral representations:
\begin{align}\label{C1}
\Sigma_{\pm}(z) = \frac{g^{2}}{(2\pi)^{2}}\iint dk_{+}dk_{-}\frac{\mathcal{O}_{\pm}(k_{+},k_{-},n_{\blacklozenge})}{z+4J\cos k_{+}\cos k_{-}},
\end{align}
where the notations $k_{\pm}\equiv\frac{1}{2}(k_{x}\pm k_{y})$ are introduced, and the numerators $\mathcal{O}_{\pm}(k_{+},k_{-},n_{\blacklozenge}) $ are given by
\begin{widetext}
\begin{align}\label{C2}
\mathcal{O}_{\pm}(k_{+},k_{-},n_{\blacklozenge}) = & 4[1+\cos(2k_{+}n_{\blacklozenge})\cos(2k_{-}n_{\blacklozenge})+\cos(2k_{+}n_{\blacklozenge})+\cos(2k_{-}n_{\blacklozenge})]\pm 4\cos(k_{+}n_{\blacklozenge})\cos(k_{-}n_{\blacklozenge})(1+2\cos[2k_{-}n_{\blacklozenge}])\nonumber\\
& \pm\left ( \cos(3k_{+}n_{\blacklozenge})+\cos(k_{-}n_{\blacklozenge})\right) ^{2}.
\end{align}
\end{widetext}
After several algebra, the self-energies $\Sigma_{\pm}(z)$ in Eq.(\ref{C1}) can be rewritten as
\begin{widetext}
\begin{align}\label{C3}
\Sigma_{\pm}(z) = \frac{4g^{2}}{(2\pi)^{2}}\iint dk_{+}dk_{-}\frac{16\cos^{2}(k_{+}n_{\blacklozenge})\cos^{2}(k_{-}n_{\blacklozenge})\pm16\cos^{3}(k_{+}n_{\blacklozenge})\cos^{3}(k_{-}n_{\blacklozenge})}{z+4J\cos k_{+}\cos k_{-}}.
\end{align}
\end{widetext}
\begin{figure}
  \centering
  \includegraphics[width=8.7cm]{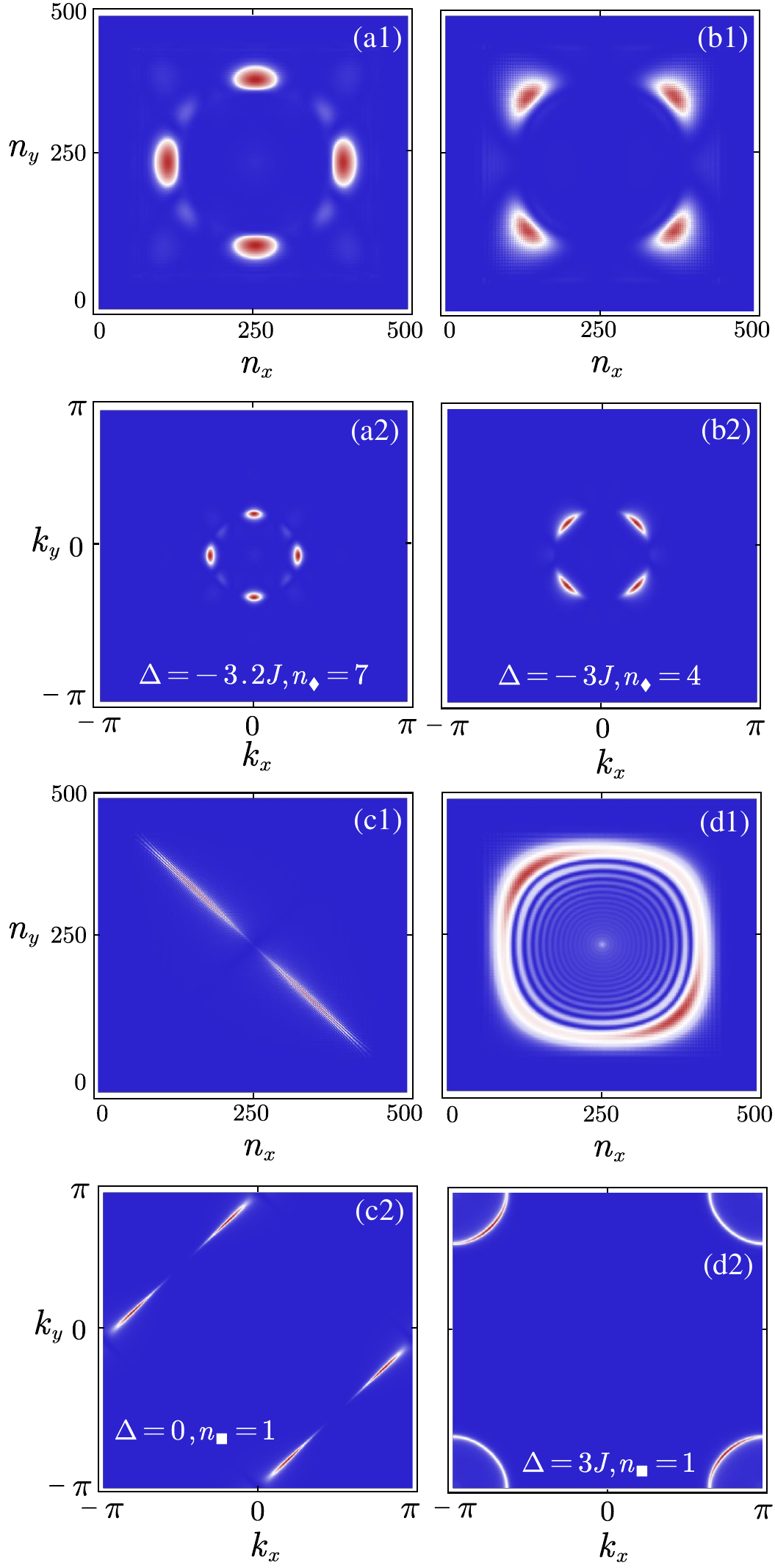}
  \caption{Bath population distribution $|C_{\boldsymbol{n}}^{\pm}(t)|$ in real space [(a1)-(d1)] and $|C_{\boldsymbol{k}}^{\pm}(t)|$ in momenta space [(a2)-(d2)] at time $tJ=100$. The simulations in panels (a) and (b) [(c) and (d)] are performed by considering two DGAs (SGAs) coupled to the 2D bath with physical parameters: (a) $n_{\blacklozenge}=7,\Delta=-3.2J,\ket{\psi(0)}=\ket{-}$; (b) $n_{\blacklozenge}=4,\Delta=-3J,\ket{\psi(0)}=\ket{-}$; $\big[$(c) $n_{\protect\scalebox{0.55}{$\protect\blacksquare$}}=1,\Delta=0,\ket{\psi(0)}=\ket{+}$; (d) $n_{\protect\scalebox{0.55}{$\protect\blacksquare$}}=1,\Delta=3J,\ket{\psi(0)}=\ket{+}$ $\big]$. All panels share the identical light-matter coupling strength $g=0.2J$.}\label{fig8}
\end{figure}
Given the fact that the bosonic modes $k_{x}\pm k_{y}=\pm(\mp)\pi$ are primarily populated when atoms locally couple to the bath with $\Delta =0$, we can approximate the numerators as $\mathcal{O}_{\pm}(\frac{\pm\pi}{2},\frac{\pm\pi}{2},n_{\blacklozenge})$. For odd $n_{\blacklozenge}$, this leads to the exact vanishing $\Sigma_{\pm}(z)=0$, thereby finding a bound state in the continuum. Based on this, we immediately obtain the steady-state populations
\begin{align}\label{C4}
C_{\pm}(t=\infty)=\frac{1}{1-\partial_{z}\left.\Sigma_{\pm}(z)\right|_{z=0}}=\frac{1}{1+n_{\blacklozenge}^{2}g^{2}/J^{2}}.
\end{align}
We now present a comprehensive study of the rich lattice dynamics of the two-dimensional square-like lattice when GAs are prepared in certain entangled states. By analyzing the photon emission in both real space and momentum space, we uncover intriguing radiation patterns due to the intricate interference effects.

As shown in Fig.\,\ref{fig8}, we present the bath population at time $tJ=100$, obtained by numerically integrating the exact dynamics govern by Eq.~(\ref{B2}) for different atomic configurations. We focus on two representative cases: (i) a DGA pair with $n_{\blacklozenge}=7,\Delta=-3.2J$ [Fig.\,\ref{fig8}(a)] and $n_{\blacklozenge}=4,\Delta=-3J$ [Fig.\,\ref{fig8}(b)]; (ii) a SGA pair with $n_{\protect\scalebox{0.55}{$\protect\blacksquare$}}=1,\Delta=0$ [Fig.\,\ref{fig8}(c)] and $n_{\protect\scalebox{0.55}{$\protect\blacksquare$}}=1,\Delta=3J$ [Fig.\,\ref{fig8}(d)]. These exotic radiation phenomena stem from the delicate interference of photons emitted by the coupled giant atom pairs.  Specially, for a pair of DGAs, we observe directional propagation of emission parallel to the principal axes of the 2D square lattice, forming spatially confined photon wave packets with sharply localized momentum distributions at $(0, \pm k),(\pm k,0)$ [Fig.~\ref{fig8}(a)]. Upon tuning the atomic geometry to modify the photonic interference, the wave packets undergo a $\pi/4$ rotation in propagation-direction, now channeled along the two diagonals, and the populated momentum modes are rotated as well [Fig.~\ref{fig8}(b)]. This contrasts distinctly with the small atom case, where a homogeneous ring distribution of photon momentum centered at the lower band edge $(k_{x} = k_{y}= 0)$ implies the absence of directional emission. Furthermore, the SGAs exhibit quasi-1D radiation that propagates exclusively along one diagonal direction [Fig.~\ref{fig8}(c)], where the modes along the lines $k_{x} - k_{y} = \pm \pi$ (mod $2\pi$, excluding $|k_{x}| = |k_{y}|$) are populated primarily in the case of $\Delta=0$. Notably, this quasi-1D radiation can also be achieved in a single SGA through coupling phase engineering\,\cite{PhysRevLett.122.203603}. Finally, we capture ripple-like emission patterns with momentum modes localized in four quadrant rings centered at upper band edge $(|k_{x}| = |k_{y}| = \pi)$ [Fig.~\ref{fig8}(d)], demonstrating rich emission physics therein.

\begin{figure}
  \centering
  \includegraphics[width=8.6cm]{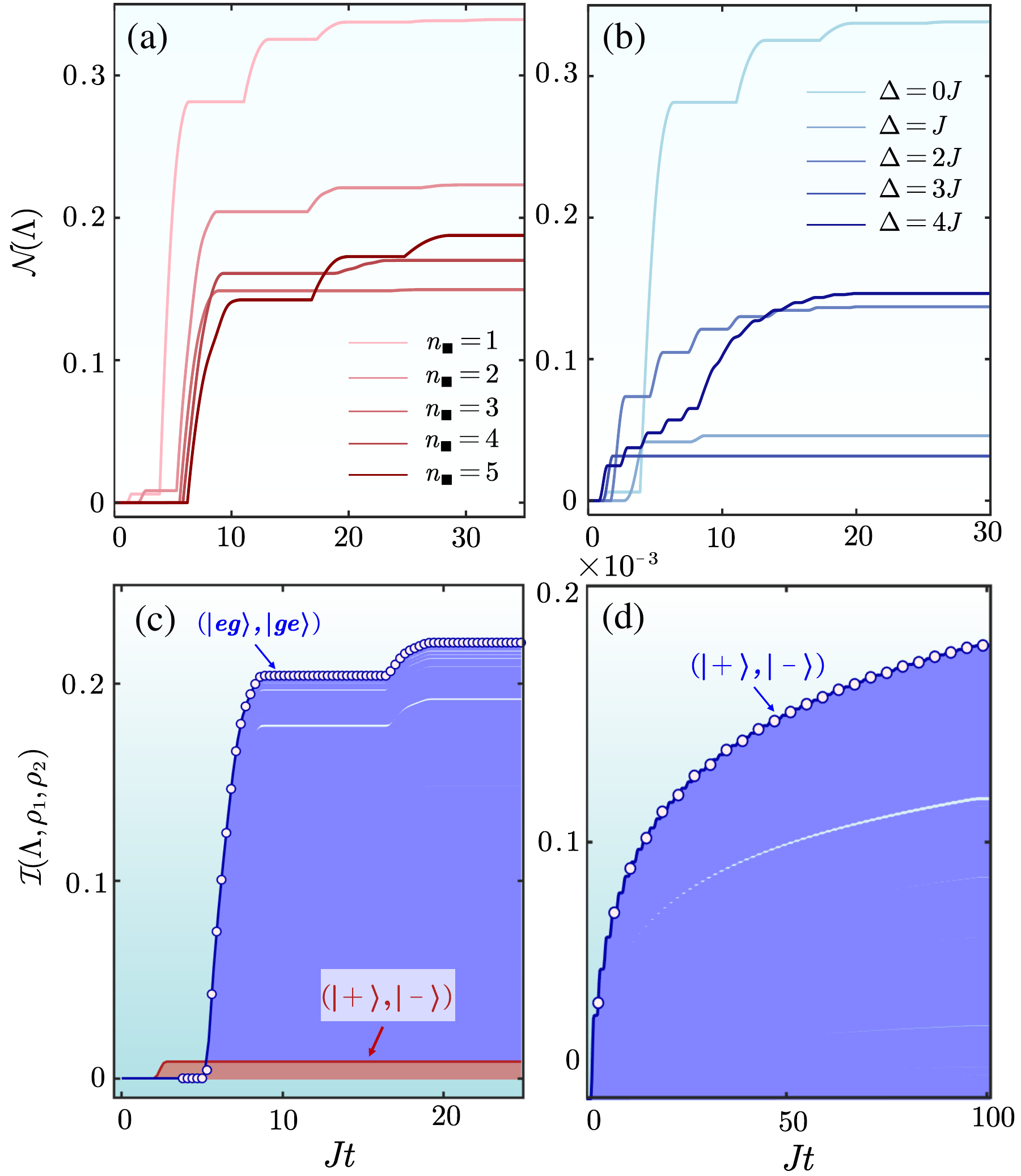}
  \caption{(a) and (b) are the measured non-Markovianity $\mathcal{N}(\Lambda)$ as a function of $Jt$ for a pair of SGAs with different atomic sizes and detuning values. Each measurement is performed by maximizing the integrals $\mathcal{I}(\Lambda,\rho_{1},\rho_{2})$ over an ensemble of $10,000$ randomly sampled initial state pairs $(\rho_{1},\rho_{2})$. Panels (c) and (d) respectively plot the integrated non-Markovian measure $\mathcal{I}(\Lambda,\rho_{1},\rho_{2})$ as a function of $Jt$ for SGAs ($n_{\protect\scalebox{0.55}{$\protect\blacksquare$}}=2$) and DGAs ($n_{\blacklozenge}=1$) with $\Delta=0$. The evolutionary trajectories maximizing the non-Markovianity $\mathcal{N}(\Lambda)$ are explicitly identified. All panels share the identical light-matter coupling strength $g=0.25J$.}\label{fig9}
\end{figure}

\section{NON-MARKOVIANITY MEASURE VIA TRACE DISTANCE}\label{D}
In this section, we rigorously quantify the non-Markovian character of the quantum system by analyzing the dynamical behavior of the trace distance between pairs of quantum states. Specifically, the revival of trace distance over time serves as a direct indicator of non-Markovianity, as it reflects the backflow of information from the environment to the system, i.e., a hallmark of memory effects. We adopt the measure proposed by Breuer-Laine-Piilo (BLP criterion)\,\cite{SMPhysRevLett.103.210401, SMPhysRevA.81.062115}, where the positive time derivative of the trace distance determines the non-Markovian regime. Thus, the degree of non-Markovianity $\mathcal{N}(\Lambda)$ can be quantified by integrating the positive contributions of the time derivative of the trace distance $D(\Lambda_{t}\rho_{1},\Lambda_{t}\rho_{2})\equiv\Vert \Lambda_{t}\rho_{1}-\Lambda_{t}\rho_{2}\Vert_{1}/2$ over the entire evolution
\begin{align}\label{D1}
\mathcal{N}(\Lambda) = \mathop{\text{max}}\limits_{\rho_{1}(0),\rho_{2}(0)}\int_{\partial_{t}D>0}dt\frac{d D(\Lambda_{t}\rho_{1},\Lambda_{t}\rho_{2})}{dt},
\end{align}
where $\Lambda_{t}$ denotes the dynamical map, and  the maximum is taken over all pairs of initial states $(\rho_{1},\rho_{2})$.

We proceed by calculating $\mathcal{N}(\Lambda)$ for a pair of SGAs, where the evolution initiated from $10000$ random quantum state pairs $(\rho_{1},\rho_{2})$. Note that the total time evolution of the GAs together with the 2D bath is simulated via the split-operator method, applicable to arbitrary initial states of the form $\cos\theta\ket{eg} + \sin\theta e^{i\varphi}\ket{ge}$. In Fig.\,\ref{fig9}, we present the measured non-Markovianity by changing the atomic sizes ranging from $n_{\protect\scalebox{0.55}{$\protect\blacksquare$}}=1$ to $n_{\protect\scalebox{0.55}{$\protect\blacksquare$}}=5$ with fixed $\Delta=0$ [see Fig.\,\ref{fig9}\,(a)], and by changing the detuning values ranging from $\Delta=0$ to $\Delta=4J$ with fixed $n_{\protect\scalebox{0.55}{$\protect\blacksquare$}}=1$ [see Fig.\,\ref{fig9}\,(b)]. More specifically, the non-Markovianity of the system preserves causality. The onset of nonzero  $\mathcal{N}(\Lambda)$ exhibits a delayed emergence as the atomic size parameter $n_{\protect\scalebox{0.55}{$\protect\blacksquare$}}$ increases, indicating size-dependent memory effects. The dependence of non-Markovianity on atomic size and detuning exhibits non-monotonic behavior, arising from the intricate interplay between BCD and UP contributions. We also show that the maximum of $\mathcal{I}(\Lambda,\rho_{1},\rho_{2})\equiv\int_{\partial_{t}D>0}dt\frac{d D(\Lambda_{t}\rho_{1},\Lambda_{t}\rho_{2})}{dt}$ can be achieved by either a pair of maximally entangled states ($\ket{+}$ and $\ket{-}$) or a pair of separable states $\ket{eg}$ and $\ket{ge}$ [see Fig.\,\ref{fig9}\,(c)], consistent with observations in other literatures\,\cite{PhysRevA.89.012307, Dziewior}.

It is noteworthy that the non-Markovianity $\mathcal{N}(\Lambda)$ for a pair of DGAs exhibits persistent monotonic growth without saturation when $\Delta=0$ [see Fig.\,\ref{fig9}\,(d)]. The observed growth of trace distance originates from the real energy splitting between the bound states $\ket{+}$ and $\ket{-}$, rather than genuine environment-induced memory effects. While mathematically qualifying as non-Markovianity under BLP criterion, this behavior may not reflect the conventional understanding of information backflow.
%
\end{document}